\documentclass[12pt,preprint]{aastex}
\usepackage{natbib}
\usepackage{amsmath}
\usepackage{amsbsy}
\usepackage{color}
\usepackage{rotating}
\usepackage[breaklinks,colorlinks,urlcolor=blue,citecolor=blue,linkcolor=blue]{hyperref}

\usepackage{graphicx}
\slugcomment{Accepted for publication in JGR Space Physics}
\shorttitle{MODEL FOR STEALTH CMES}
\shortauthors{LYNCH ET AL.}

\begin{document}

%
%

\title{A model for stealth coronal mass ejections}
%
%

%
%



\author{B.~J.~Lynch\altaffilmark{1}, 
        S.~Masson\altaffilmark{2}, 
        Y.~Li\altaffilmark{1}, 
        C.~R.~DeVore\altaffilmark{3}, 
        J.~G.~Luhmann\altaffilmark{1},
        S.~K.~Antiochos\altaffilmark{3},
        and G.~H.~Fisher\altaffilmark{1}}

\affil{\altaffilmark{1}Space Sciences Laboratory, University of California,
        Berkeley, CA 94720, USA}
\affil{\altaffilmark{2}LESIA, Observatorie de Paris, PSL Research University,
	CNRS, Sorbonne Universit\'{e}s, UPMC Univ. Paris 06, Univ.
	Paris Diderot, Sorbonne Paris Cit\'{e}, France}
\affil{\altaffilmark{3}Heliophysics Science Division, NASA Goddard Space
        Flight Center, Greenbelt, MD 20771, USA}

\begin{abstract}

Stealth coronal mass ejections (CMEs) are events in which there are
almost no observable signatures of the CME eruption in the low
corona but often a well-resolved slow flux rope CME observed in the
coronagraph data.  We present results from a three-dimensional
numerical magnetohydrodynamics (MHD) simulation of the 1--2 June
2008 slow streamer blowout CME that \citet{Robbrecht2009} called
``the CME from nowhere.'' We model the global coronal structure
using a 1.4~MK isothermal solar wind and a low-order potential field
source surface representation of the Carrington Rotation 2070
magnetogram synoptic map. The bipolar streamer belt arcade is
energized by simple shearing flows applied in the vicinity of the
helmet streamer's polarity inversion line. The flows are large scale
and impart a shear typical of that expected from the differential
rotation.  The slow expansion of the energized helmet streamer
arcade results in the formation of a radial current sheet. The
subsequent onset of expansion-induced flare reconnection initiates
the stealth CME while gradually releasing the stored magnetic energy.
We present favorable comparisons between our simulation results and
the multiviewpoint SOHO-LASCO (Large Angle and Spectrometric Coronagraph) 
and STEREO-SECCHI (Sun Earth Connection Coronal and 
Heliospheric Investigation) coronagraph observations of the preeruption 
streamer structure and the initiation and evolution of the stealth streamer 
blowout CME.

\end{abstract}

%
%

%


%
%

\section{Introduction}
\label{sxn_intro}

Stealth coronal mass ejections (CMEs) are CME events with virtually
no identifiable surface or low corona signatures that indicate an
eruption has occurred.  The stereoscopic viewing and improvements
in heliospheric white-light imaging by STEREO Sun Earth Connection 
Coronal and Heliospheric Investigation (SECCHI) instruments, combined 
with the last solar minimum's exceptionally
low activity levels, have dramatically increased the number 
and proportion of events now classified as stealth CMEs. This is
partly due to more eruptive transients being classified as CMEs,
fewer fast CMEs during the early years of the STEREO mission and
the stereoscopic viewing being able to rule out events that would
have been previously classified as ``backside halo CMEs''
\citep{Howard2013}. Whether stealth CMEs constitute a distinct
class of CME event in their own right or merely represent a subset
of a more common or universal set of CME phenomena is still under
debate.  Stealth CMEs have a particular relevance for space weather
forecasting as potentially geoeffective interplanetary CME (ICME)
structures that lack the usual solar eruption warning signs.

\citet{Howard2013} presented a historic perspective of the stealth
CME phenomena and argue that they are, in fact, merely the subset
of slow, streamer blowout type of CMEs \citep{Sheeley2007b} that
have been seen since \emph{Skylab}.  \citet{Ma2010} conducted a
statistical study of stealth CME events and concluded their eruption
velocities in the coronagraph fields of view were consistent with---and essentially defined---the low end of the CME speed distribution
($\lesssim 300$~km~s$^{-1}$).  \citet{Pevtsov2012} examined properties
of the solar source regions for several slow- to moderate-speed CME
events that originated in ``empty" filament channels, i.e. long
polarity inversion lines in the underlying photospheric magnetic
field distribution with highly sheared coronal magnetic fields but
no discernible filament or prominence material.
\citet{DHuys2014} followed up on the study by
\citet{Ma2010} by presenting a set of $\sim$40 stealth CME events.
\citet{Alzate2016SPD} have shown that with sufficiently advanced
image-processing techniques and the extended field of view of the PROBA2
coronagraph, the vast majority of the \citet{DHuys2014} stealth CME
events do, in fact, have some low coronal signatures. These findings
confirm our main conclusion here, that stealth CMEs are not a new,
mysterious, and unknown eruptive phenomena.

\citet{Robbrecht2009} analyzed the STEREO observations of the 1--2 June 2008 
stealth CME eruption. The STEREO viewpoint geometry was
such that the CME was a near-perfect limb event in STEREO-A coronagraph
and heliospheric imager observations and was directed toward STEREO-B.
The in situ STEREO-B plasma and magnetic field observations showed
classic flux rope ICME signatures at 1 AU. Therefore, there have
been a number of detailed studies of various aspects of this CME-ICME
event that link the remote and in-situ measurements
\citep[e.g.][]{Robbrecht2009, Moestl2009, Bisi2010, Lynch2010,
Wood2010, Nieves-Chinchilla2011, Nieves-Chinchilla2012, Rodriguez2011,
Rollett2012}.

There are several competing models for the initiation of coronal
mass ejections and several competing mechanisms for the gradual
accumulation of the free magnetic energy necessary for release
during the eruption \citep[e.g., see reviews by][and references
therein]{Klimchuk2001, Forbes2006SSR, Aulanier2014IAU, Janvier2015}.
In general, regardless of the specific ideal or resistive
magnetohydrodynamic (MHD) instabilities associated with the rapid
transition to a catastrophic, run-away eruption of coronal fields
and plasma, essentially every CME model leads to magnetic reconnection
at a vertical/radial current sheet that drives the eruption in the
manner of the CSHKP eruptive flare scenario \citep{Carmichael1964,
Sturrock1966, Hirayama1974, KoppPneuman1976}.

One of the simplest ways to model the energization of a magnetic
field arcade in MHD simulations is through the gradual shearing of
the arcade foot points parallel to the polarity inversion line of
the magnetic configuration \citep{DeVore2000b}.  \citet{Mikic1994}
showed that given sufficient energization, a spherical, axisymmetric
sheared dipole arcade could create such an extended radial current
sheet so that, eventually, magnetic reconnection would facilitate the
rapid formation of a disconnected flux rope that is ejected out of
the simulation domain. \citet{Linker1995} repeated this form of
driving in an axisymmetric bipolar helmet streamer configuration
and showed the presence of a background solar wind increased the
energy of the erupting structure. A fully 3-dimensional (3-D),
spherical version of a sheared arcade eruption was presented by
\citet{Linker2003PhPl}.

In the MHD simulation results presented here, we utilize this same
basic energization mechanism---the slow gradual shearing of a 3-D
helmet streamer arcade in a simple isothermal solar wind---and
model the evolution and eruption of a slow, streamer blowout-type
CME that reproduces many of the observational properties of the
1--2 June 2008 STEREO stealth CME event. Our results strongly support
the \citet{Howard2013} conjecture that stealth CMEs are not
fundamentally different than most slow CME eruptions but just
represent the lowest-energy range of the CME distribution.

The paper is organized as follows. In section~\ref{sxn.setup} we
describe the numerical methods and the MHD simulation setup,
including the computational grid and boundary conditions, initial
potential field source surface configuration derived from the
observed magnetograph synoptic map, the isothermal wind model and
its relaxation to steady state. We then compare the global,
quasi-steady state model streamer structure to multispacecraft
coronagraph observations. In section~\ref{sxn.cme} we present the
simulation results. First, we describe the shear flows used to
energize the bipolar helmet streamer arcade, then we present the
global energy evolution of the system and examine the stealth CME
initiation, and finally we examine the magnetic field structure and
evolution of the CME flux rope as it forms and propagates, including
its dynamics through the coronagraph field of view. In
section~\ref{sxn.disc} we discuss the implications of our modeling
results for CME initiation and global coronal evolution and for
space weather forecasting. In section~\ref{sxn.concl} we summarize
our results and present the conclusions.

\section{Numerical Simulation Setup}
\label{sxn.setup}
\subsection{ARMS Model Description}
\label{ssxn.arms}

The Adaptively Refined MHD Solver (ARMS) \citep{DeVore2008}
calculates solutions to the 3-D nonlinear, time-dependent MHD equations
that describe the evolution and transport of density, momentum, and
energy throughout the plasma, and the evolution of the magnetic
field and electric currents using a finite-volume, multidimensional
flux-corrected transport numerical scheme \citep{DeVore1991}.  The
ARMS code is fully integrated with the adaptive mesh toolkit PARAMESH
\citep{MacNeice2000} to handle solution-adaptive grid refinement
and support efficient multiprocessor parallelization.
ARMS has been used to perform a wide variety of
numerical simulations of dynamic phenomena in the solar atmosphere,
including 3-D magnetic breakout CME initiation \citep{DeVore2008,
Lynch2008} and ejecta propagation through the low corona
\citep{Lynch2009}.

For our simulation, we use ARMS to solve the ideal MHD equations
in spherical coordinates,
\begin{equation}
    \frac{\partial \rho}{\partial t} + \nabla \cdot \left( \rho
    \boldsymbol{v} \right) = 0 ,
    \label{eq1}
\end{equation}
\begin{equation}
    \frac{\partial}{\partial t} \left( \rho \boldsymbol{v} \right)
    + \nabla \cdot \left( \rho \boldsymbol{v} \boldsymbol{v} \right)
    + \nabla P = \frac{1}{4\pi}\left( \nabla \times \boldsymbol{B}
    \right) \times \boldsymbol{B} - \rho \boldsymbol{g} ,
    \label{eq2}
\end{equation}
\begin{equation}
    \frac{\partial \boldsymbol{B}}{\partial t} = \nabla \times
    \left( \boldsymbol{v} \times \boldsymbol{B} \right),
    \label{eq4}
\end{equation}
where all the variables retain their usual meaning, solar
gravity is $\boldsymbol{g} = g_\odot (r/R_\odot)^{-2} \boldsymbol{\hat{r}}$
with $g_\odot = 2.75\times10^4$~cm~s$^{-2}$, and we use the ideal
gas law $P=2(\rho/m_p)k_BT$.  However, given the isothermal model
used in the construction of our background solar wind, we do not
solve an internal energy or temperature equation. The plasma
temperature remains uniform throughout the domain for the duration
of the simulation.  Since coronal plasma is highly conductive and
the dominant heat flux is carried by fast-moving electrons, the
isothermal assumption of an ``infinite conductivity'' is a reasonable
approximation.

Additionally, while there is no explicit magnetic resistivity in
the equations of ideal MHD, necessary and stabilizing numerical
diffusion terms introduce an effective resistivity on very small
spatial scales, i.e., the size of the grid. In this way, magnetic
reconnection can occur when current sheet features and the associated
gradients of field reversals have been distorted and compressed to
the local grid resolution scale.

The spherical computational domain uses logarithmic grid spacing
in $r$ and uniform grid spacing in $\theta, \phi$. The domain extends
from $r \in \left[1R_\odot, 30R_\odot\right]$, $\theta \in
\left[0.0625\pi,0.9375\pi\right]$, and $\phi \in \left[-\pi,
\pi\right]$.  The initial grid consists of $8 \times 7 \times 16$
blocks with 8$^3$ grid cells per block. There are 2 additional
levels of static grid refinement: the level 2 refinement covers $r
\le 3.58R_\odot$ for $\phi \in \left[-0.1875\pi, \pi\right]$ and
all $\theta$; the level 3 refinement is over $r \le 3.58R_\odot$
for $\phi \in \left[-\pi, -0.1875\pi\right]$ for all $\theta$ and
includes an additional spherical wedge through $30R_\odot$ from
$\theta \in \left[0.375\pi, 0.750\pi\right]$ and $\phi \in
\left[-0.9375\pi, -0.3125\pi\right]$, encompassing the volume above
the streamer blowout eruption.

\begin{figure}
\includegraphics[width=6.5in]{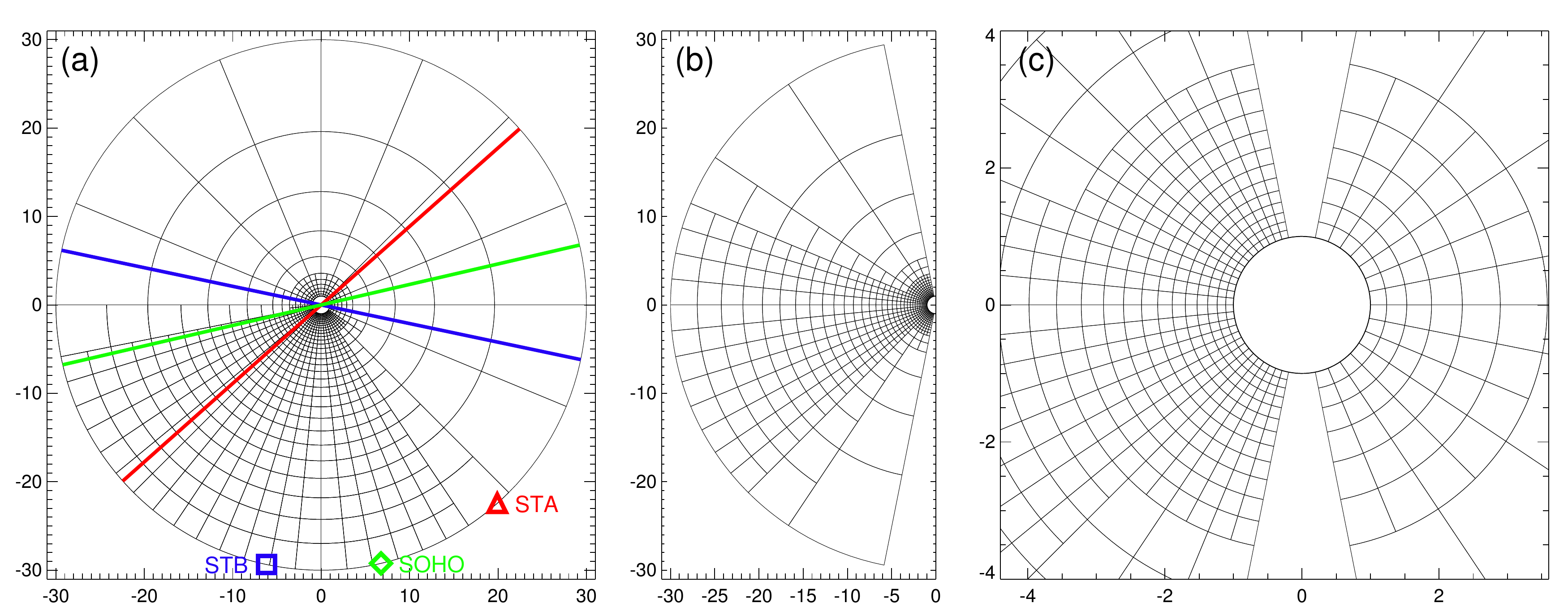}
\caption{The 3-D spherical $\log{r}$ computational block structure.
	 (a) The $r$-$\phi$ plane with the relative heliographic longitude
	 separation of the STEREO and SOHO spacecraft on 2 June 2008 
	 (labeled plot symbols) as well as their respective
	 plane-of-the-sky orientations (STA--red; SOHO--green;
	 STB--blue). (b) The $r$-$\theta$ plane corresponding to the
	 east limb viewpoint of STEREO-A. (c) A zoomed-in view of
	 the static block refinement near the Sun in Figure~\ref{f1}b.
         \label{f1}
}
\end{figure}

Figure~\ref{f1} shows the computational block structure. Figure~\ref{f1}a
plots the equatorial $(r, \phi)$ plane. The angular positions of
the STEREO-B (blue square; $\phi_{\rm STB}=-101.9^\circ$), SOHO
(green diamond; $\phi_{\rm SOHO}=-77.0^\circ$), and STEREO-A (red
triangle; $\phi_{\rm STA}=-48.4^\circ$) spacecraft on 2 June 2008
are also shown (converted from Carrington Longitude to the ARMS
coordinate system). The STB, SOHO, and STA planes of sky are plotted
as the solid lines. Figure~\ref{f1}b plots the block structure in the $(r,
\theta)$ plane corresponding to the east limb of STA plane of sky.
Figure~\ref{f1}c plots the zoomed-in view of inner corona for the STA
plane of sky.  This grid structure yields a modest resolution of
$\sim$$1.6 \times 10^4$ leaf blocks and $\sim$$8.2 \times 10^6$
computational grid cells. The highest refinement region on the east
limb corresponds to a global grid size of $256 \times 224 \times
512$ in $({\log r}, \theta, \phi)$, a radial resolution of
0.0134$R_\odot$ at $r = 1R_\odot$, and an angular resolution of
$0.703^\circ \times 0.703^\circ$.

The simulation's boundary conditions are the following.  The left
and right $\phi$ boundaries are periodic.  The left and right
$\theta$ boundaries are closed, which sets all fluxes across the
boundary to zero.  The left (bottom) and right (top) radial boundary
conditions are set to allow the implementation of our solar wind
solution described in section~\ref{ssxn.sw}.  The guard cells of
the radial boundaries are filled symmetrically, and both the inner
and outer boundaries are open to fluxes of all quantities. The inner
$r$ boundary guard cells are filled with values of the initial
atmospheric density, pressure, and temperature at every time step,
but the velocities in the guard cells are set to zero. This creates
a passive absorbing layer in which the normal flows across the
boundary arise via the averaging to obtain the value at the cell
face. These flows compensate for the differences in density/pressure
that develop during the evolution of the system.  This absorbing
layer also acts as an effective mass source which is necessary to
maintain a quasi-steady state solar wind outflow.  The outer
$r$ boundary guard cells are filled with the solution at the edge
of the domain multiplied by a factor corresponding to the radial
dependence of the initial atmospheric profile. The outer radial
boundary allows flow through in the same manner as the inner boundary.
Since the quasi-steady state solar wind outflow is supersonic and
super-Alfv\'{e}nic by 30$R_\odot$, the influence of the outer
boundary conditions on the simulation is minimal.

\subsection{Initial Magnetic Field}
\label{ssxn.init}

We initialize the ARMS simulation magnetic field with the potential
field source surface (PFSS) reconstruction \citep[e.g.,][]{Wang1992,
Luhmann1998} of the SOHO-MDI \citep{Scherrer1995} polar-corrected
synoptic map data \citep{Liu2007} for Carrington Rotation 2070 with
the source surface set to 2.5$R_\odot$.  \citet{Robbrecht2009} and
\citet{Lynch2010} estimated the 1--2 June 2008 stealth CME source
region as a large swath of the helmet streamer belt, so here we can
use a low-order PFSS $(\ell_{\rm max}=9)$ extrapolation and still
capture the overall structure of the helmet streamer.

\begin{figure}
\includegraphics[width=6.5in]{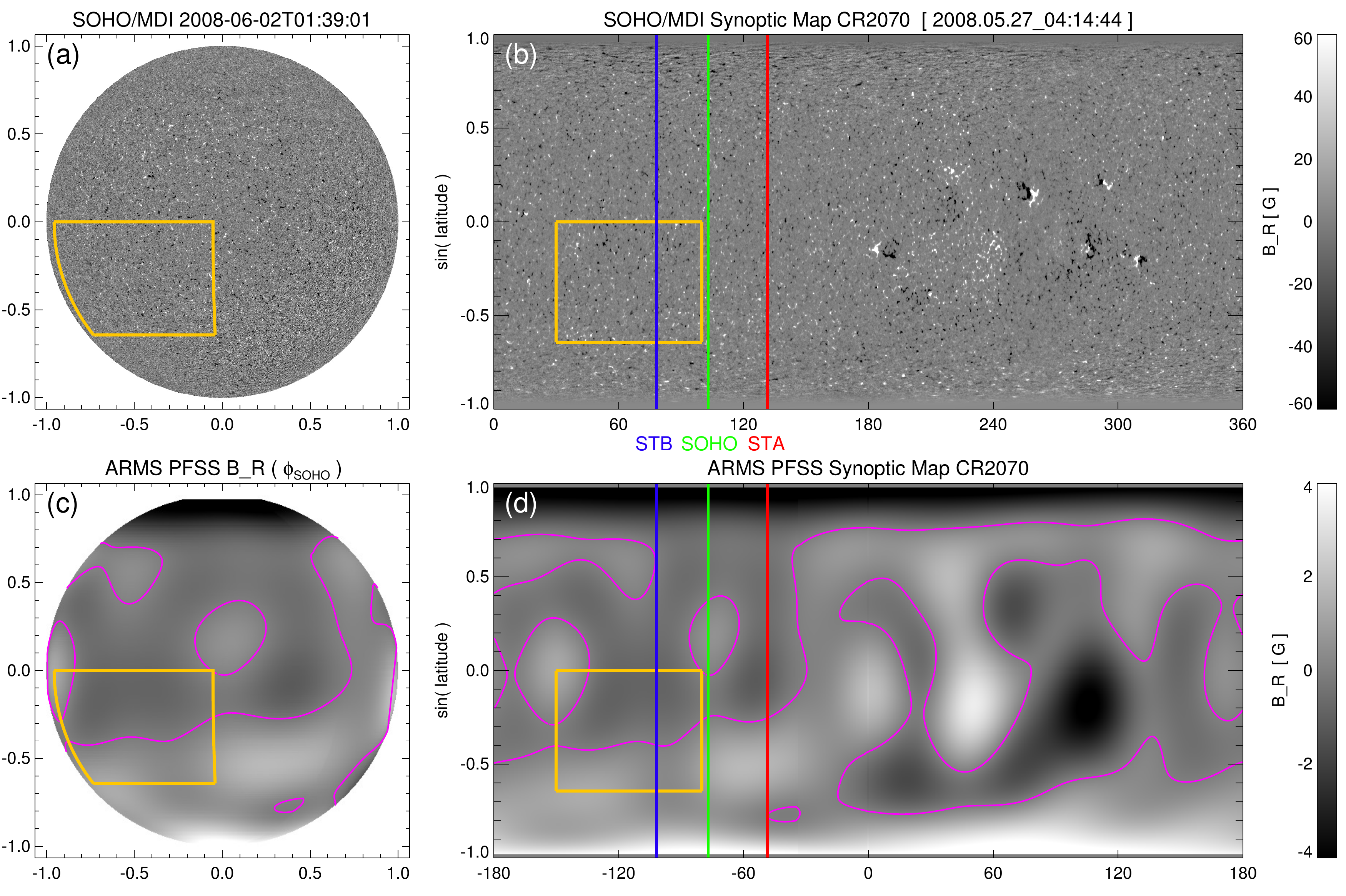}
\caption{The observed photospheric and model coronal fields. (a)
	 SOHO-MDI magnetogram on 2 June 2008. The \citet{Robbrecht2009}
	 estimate of the stealth CME source region is shown as the
	 yellow region. (b) MDI synoptic map for CR2070. The
	 respective spacecraft longitudes are shown as vertical
	 lines (STA--red; SOHO--green; STB--blue).  (c) ARMS
	 simulation radial field distribution from SOHO/L1 viewpoint.
	 The $B_r=0$ polarity inversion lines are plotted in magenta.
	 (d) Low-order PFSS representation of the CR2070 synoptic
	 map.
         \label{f2}
}
\end{figure}

Figure~\ref{f2} plots the SOHO/MDI data and the initial $r=R_\odot$
magnetic field used in the MHD simulation.  Figure~\ref{f2}a shows the MDI
full disk magnetogram at 01:39UT on 2 June 2008 and Figure~\ref{f2}b plots
the CR2070 synoptic map. Here the CME source region estimate by
\citet{Robbrecht2009} is indicated in the orange box and the positions
of STEREO-B, SOHO, and STEREO-A are shown as blue, green, and red
vertical lines at their respective Carrington longitudes.  Figure~\ref{f2}c 
plots the full disk SOHO view of the ARMS radial magnetic field
and panel (d) plots the PFSS representation of the simulation's
$B_r(R_\odot, \theta,\phi)$ synoptic map. The magenta contours
represent the polarity inversion lines in the PFSS field. The
magnetic configuration of the 1--2 June 2008 CME source region becomes
apparent as the large scale polarity inversion line corresponding
to the main helmet streamer belt.

\subsection{Isothermal Solar Wind Solution}
\label{ssxn.sw}

The solar wind is initialized in ARMS by first solving the
one-dimensional \citet{Parker1958} equation for an isothermal coronal
atmosphere,
\begin{equation}
    \frac{v_{\rm sw}^2}{c_0^2} - \ln\left( \frac{v_{\rm sw}^2}{c_0^2}
    \right) = -3 + 4 \ln\left( \frac{r}{r_c} \right) + 4 \frac{r_c}{r},
\end{equation}
where the base number density, pressure, and temperature
are $n_0 = \rho_0/m_p = 1.29\times10^8$~cm$^{-3}$, $P_0 =
0.05$~dyn~cm$^{-2}$, and $T_0 = 1.40\times10^6$~K, respectively.
Here, $c_0 = (2 k_B T_0 / m_p)^{1/2} = 152$~km~s$^{-1}$ is the
thermal velocity at $T_0$, the location of the critical point is
$r_c = G M_\odot / 2 c_0^2 = 4.10R_\odot$.  This yields a solar
wind speed at the outer boundary of $v_{\rm sw}(30R_\odot) \sim
450$~km~s$^{-1}$.

\begin{figure}
\includegraphics[width=6.5in]{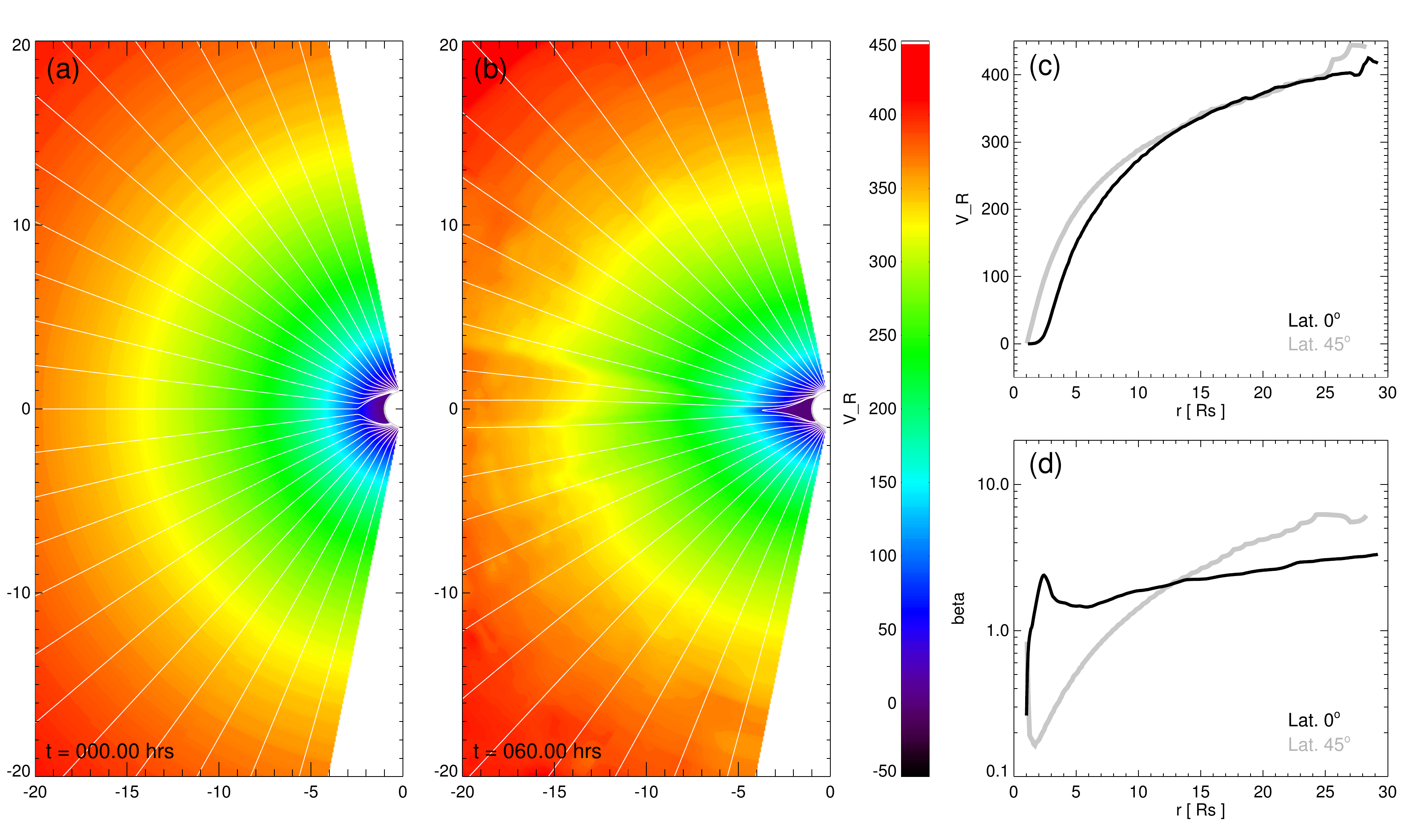}
\caption{The relaxation of the isothermal solar wind solution. (a)
	 The initial Parker $v_{\rm sw}(r)$ solution with representative
	 field lines illustrating the initial PFSS magnetic field
	 solution for the STA east limb viewpoint. (b) The same
	 quantities at the end of the relaxation phase ($t=60$~h).
	 (c) Two radial cuts through the $V_r(r,\theta)$ simulation
	 data shown in Figure~\ref{f3}b, one at the equator (black) going through
	 the helmet streamer, and the other at +45$^\circ$ (gray)
	 sampling the polar coronal hole. (d) Radial cuts for plasma
	 $\beta = 8\pi P /B^2$. An animation of the solar wind
	 relaxation (Figures~\ref{f3}a and \ref{f3}b) is available as supporting 
	 information of the article.
         \label{f3}
}
\end{figure}

At time $t=0$~s we impose this Parker $v_{\rm sw}(r)$ profile and
use it to set the initial mass density profile $\rho(r)$ from the
steady mass-flux condition ($\rho v_{\rm sw} r^2 =$~constant)
throughout the computational domain. The $v_{\rm sw}(r)$ profile
is shown as the contour plane in Figure~\ref{f3}a alongside
representative magnetic field lines of the initial PFSS magnetic
field model.  We then let the system relax for $t=2.16\times10^5$~s
(60~h) and Figure~\ref{f3}b shows a snapshot of the magnetic field and
radial velocity at the end of the relaxation phase (this figure is
available as an animation in the supporting information of this article).
The initial discontinuities in the magnetic field at the source
surface ($r=2.5R_\odot$) propagate outwards and eventually through
the outer boundary.  The Figure~\ref{f3} animation shows that the
dragging of some of the closed streamer belt flux by the solar wind
flow sets up the condition for the transverse pressure from the
open fields to push in behind the expanding streamer belt structure
forming the elongated current sheet.  Eventually, the numerical
diffusion allows magnetic reconnection between the elongated streamer
belt field lines and gives the system the opportunity to adjust the
amount of open flux relative to the new pressure balance associated
with the background solar wind flows, as seen in Figure~\ref{f3}b.  The
inner boundary mass source allows material to accumulate in the
closed field regions and sets up steady radial flows along open
field lines.

Our isothermal solar wind never actually achieves a true ``steady state''
stationary outflow.  We see a continuous, dynamic situation of
small-scale opening and closing at the Y point of the streamer belt
and in the heliospheric current sheet.  In simulations with much
higher resolution \citep[e.g., see][]{Allred2015}, interchange
reconnection and the intermittent generation of magnetic islands
in the heliospheric current sheet are likely one of the sources of
the well-known ``streamer blob'' intensity enhancements that have
been shown to passively advect with the slow solar wind
\citep{Sheeley1997,Wang2000}.

\subsection{Global Preeruption Coronal Structure}
\label{ssxn.obs.pre}

We have constructed synthetic total brightness images $I(t)$ from
the ARMS simulation data by building 2-D plane-of-sky images  where
each pixel corresponds to a line-of-sight integration of white-light
intensity associated with Thomson scattering from the 3-D density
data \citep[as in][]{Vourlidas2013}. The white-light scattering is
calculated with a Fortran implementation of the SolarSoft routine
{\tt{eltheory.pro}}, where the intensity contribution from the
simulation density along the line of sight is determined by the
observing geometry \citep[][]{Billings1966, Vourlidas2006}.  In
order to mimic the coronagraph analysis procedures, we construct
``ratio images'' from the white-light intensity images
\citep[see][]{Sheeley1997}. The $I(t)/I(0)$ ratio compensates for
the (spherically symmetric, background) density profile's radial
dependence and thus highlights the extended streamer structures.
The ratio images are equivalent to performing background subtraction
for logarithmic intensity scales.

\begin{figure}
\includegraphics[width=6.5in]{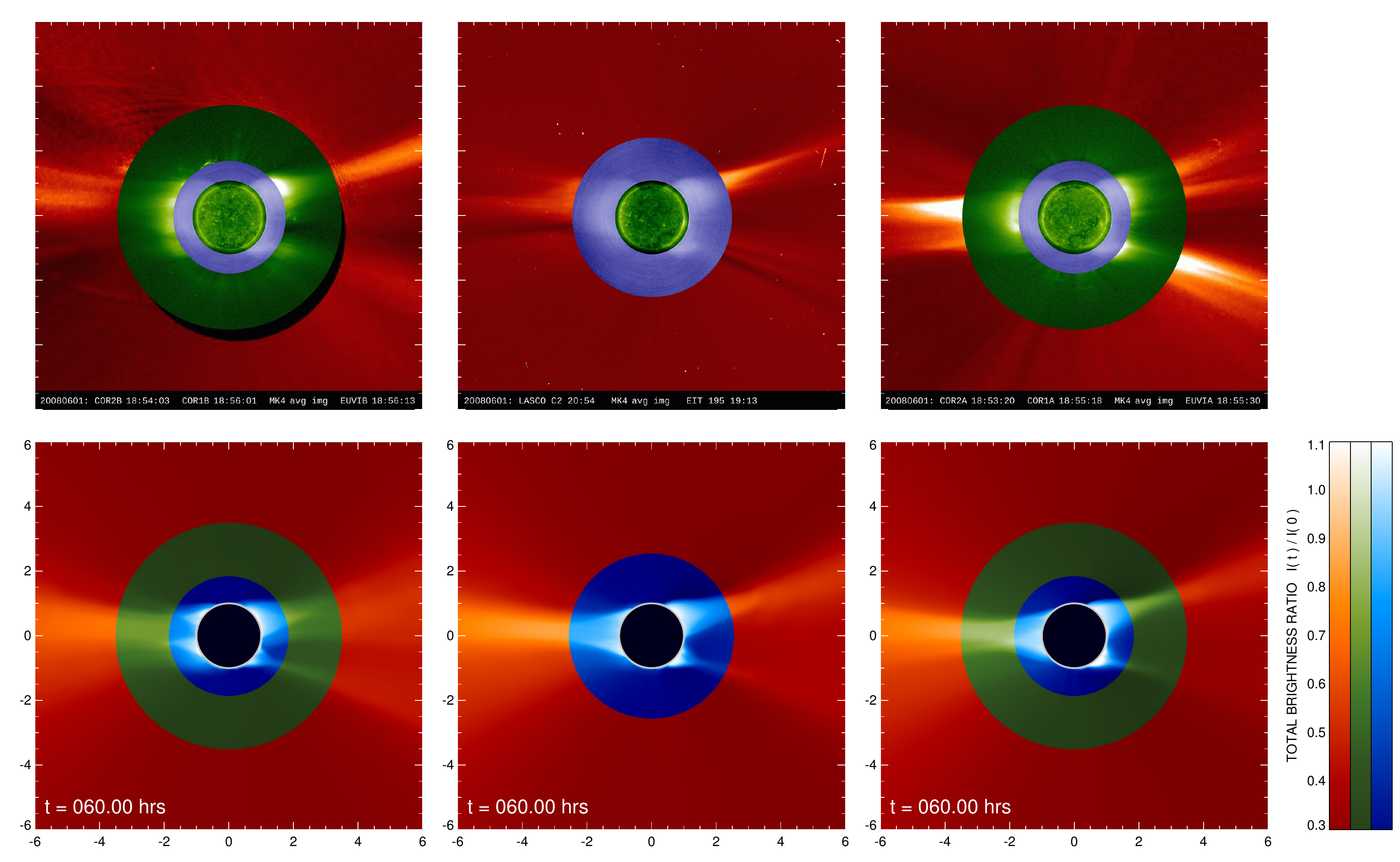}
\caption{Multispacecraft coronagraph observations and model results
	 of the preeruption streamer structure. (top row) (left) STB, (middle)
	 SOHO, and (right) STA observations supplemented
	 with averaged MK4 data from 1 June 2008. (bottom row) Synthetic
	 white-light emission ratio images of the steady state
	 coronal streamer structure at the end of the relaxation
	 phase ($t=60$~h) from the STB, SOHO, and STA viewpoints.
         \label{f5}
}
\end{figure}

Figure~\ref{f5} (top row) shows the preeruption coronal structure
as observed by multiple coronagraph instruments from the STB (left), 
SOHO (middle), and STA (right) viewpoints. Each
of the STEREO viewpoints includes observations from the COR1 and COR2
cameras from the SECCHI suite and are supplemented by the
averaged MK4 coronagraph data on
1 June 2008 from the Mauna Loa Solar Observatory. The
MK4 data are time shifted by approximately 2 days with respect to
each of the STEREO image composites ($\sim$26$^\circ$ in Carrington
longitude) but still show reasonable agreement with the extended
COR1 and COR2 large-scale helmet streamer structures. The SOHO
viewpoint data come from the SOHO/Large Angle Spectrometric Coronagraph (LASCO) C2 instrument and are
likewise supplemented by the MK4 data. Here the 195 \AA\ EUV emission
over the disk is also shown from the STEREO EUVI and SOHO EIT
instruments.  Figure~\ref{f5} (bottom row) plots the white-light
intensity ratio images from the $(\phi_{\rm STB}, \phi_{\rm SOHO},
\phi_{\rm STA})$ perspectives at $t=60$~h with color scales
chosen to correspond to the different coronagraph observations used
above.  For each of the three spacecraft perspectives, we capture
the large-scale coronal streamer and pseudostreamer structures seen
in the observations.

\section{MHD Simulation Results}
\label{sxn.cme}


\subsection{Shearing the Bipolar Helmet Streamer Arcade With Boundary Flows}
\label{ssxn.shear}

The helmet streamer is energized via ideal foot point shearing flows
imposed at the $r=R_\odot$ boundary, roughly parallel to the polarity inversion line (PIL)
in the CME source region. Here we define
\begin{equation}
    \boldsymbol{V}_{\rm shear}(\theta,\phi,t) = V_0 \left[
    \sum_{i=1}^{4} \Theta^{i}(\theta) \Phi^{i}(\phi) \right] T(t)
    \boldsymbol{\hat{\phi}} .
    \label{e.shearprofile}
\end{equation}
The separable spatial dependence functions are of the form
\begin{equation}
    \Theta^{i}(\theta) = \sin{ \left[ 2\pi k^{i}_\theta \frac{(\theta
    - \theta^{i}_C)}{(\theta^{i}_L - \theta^{i}_R)} \right] } \; ,
    \label{e.theta}
\end{equation}
\begin{equation}
    \Phi^{i}(\phi) =  \frac{1}{2}-\frac{1}{2}\cos{\left[ 2 \pi
    k^{i}_\phi \frac{(\phi-\phi^i_C)}{(\phi^i_L - \phi^i_R)} \right]},
    \label{e.phi}
\end{equation}
and the temporal dependence is 
\begin{equation}
    {\rm T}(t) = \frac{1}{2}-\frac{1}{2}\cos{\left[ 2\pi \frac{(t-60)}{20}
    \right]}.
    \label{e.time}
\end{equation}
Each of the separable functions above is defined only for
$\theta \in [\theta^i_L, \theta^i_R]$, $\phi \in [\phi^i_L, \phi^i_R]$
with respective phase shifts ($\theta^i_C$,
$\phi^i_C$), for $t \in \left[ 60, 80 \right]$~h and are zero
elsewhere.  Table~\ref{table.one} lists each of the specific
parameters for the four flow fields $(i=1,4)$ that make up the
composite shearing profile.  We have chosen $V_0 = 30$~km~s$^{-1}$
so that the spatial and temporal average of the composite profile
yields a mean velocity magnitude of $\sim$3.50~km~s$^{-1}$ in each
magnetic polarity over the 20~h duration.  Although the maximum
shearing velocity $V_0$ is roughly 30 times the magnitude of typical
photospheric velocities, it is less than both the sound speed
$(V_0/c_0 \sim 20\%)$ and the Alfv\'{e}n speed $(V_0/\langle V_A
\rangle \sim 15\%)$ in the helmet streamer region near the lower
radial boundary. Thus, the system evolves quasi-statically.

We generate a total foot point displacement of $\approx 2.6 \times
10^5$~km ($\sim 0.37 R_\odot$) in each polarity
over the 20~h temporal duration of the applied shearing flows.
This corresponds to roughly the same foot point displacement that
would occur over 15~days of differential rotation using the
\citet{Lynch2010} estimate of 0.20~km~s$^{-1}$ in the $\phi$-direction
over the 40$^\circ$ latitudinal extent of the source region.

%

\begin{table}
\begin{center}
\caption{Parameters for the Spatial Extent of the Surface Shearing
	 Profile Given by Equation~(\ref{e.shearprofile}).
	 \label{table.one}}
\begin{tabular}{rrrrrrrrr}
\tableline\tableline
   $i$ &  $k_\theta$ & $\theta_L$ (deg)  & $\theta_R$ (deg)
   & $\theta_C$ (deg) & $k_\phi$ & $\phi_L$ (deg) &
   $\phi_R$ (deg) & $\phi_C$ (deg)\\
\tableline
1  &  0.25  &  94.0  & 108.25 &  94.0  & 1.0 & -130.0 & -40.0 &  -85.0 \\
2  & -0.25  & 108.25 & 111.75 & 111.75 & 1.0 & -130.0 & -40.0 &  -85.0 \\
3  &  0.25  & 113.75 & 117.25 & 113.75 & 1.0 & -150.0 & -60.0 & -105.0 \\
4  & -0.25  & 117.25 & 129.75 & 129.75 & 1.0 & -150.0 & -60.0 & -105.0 \\
\tableline
\end{tabular}
\end{center}
\end{table}

\begin{figure}
\center
\includegraphics[width=4.25in]{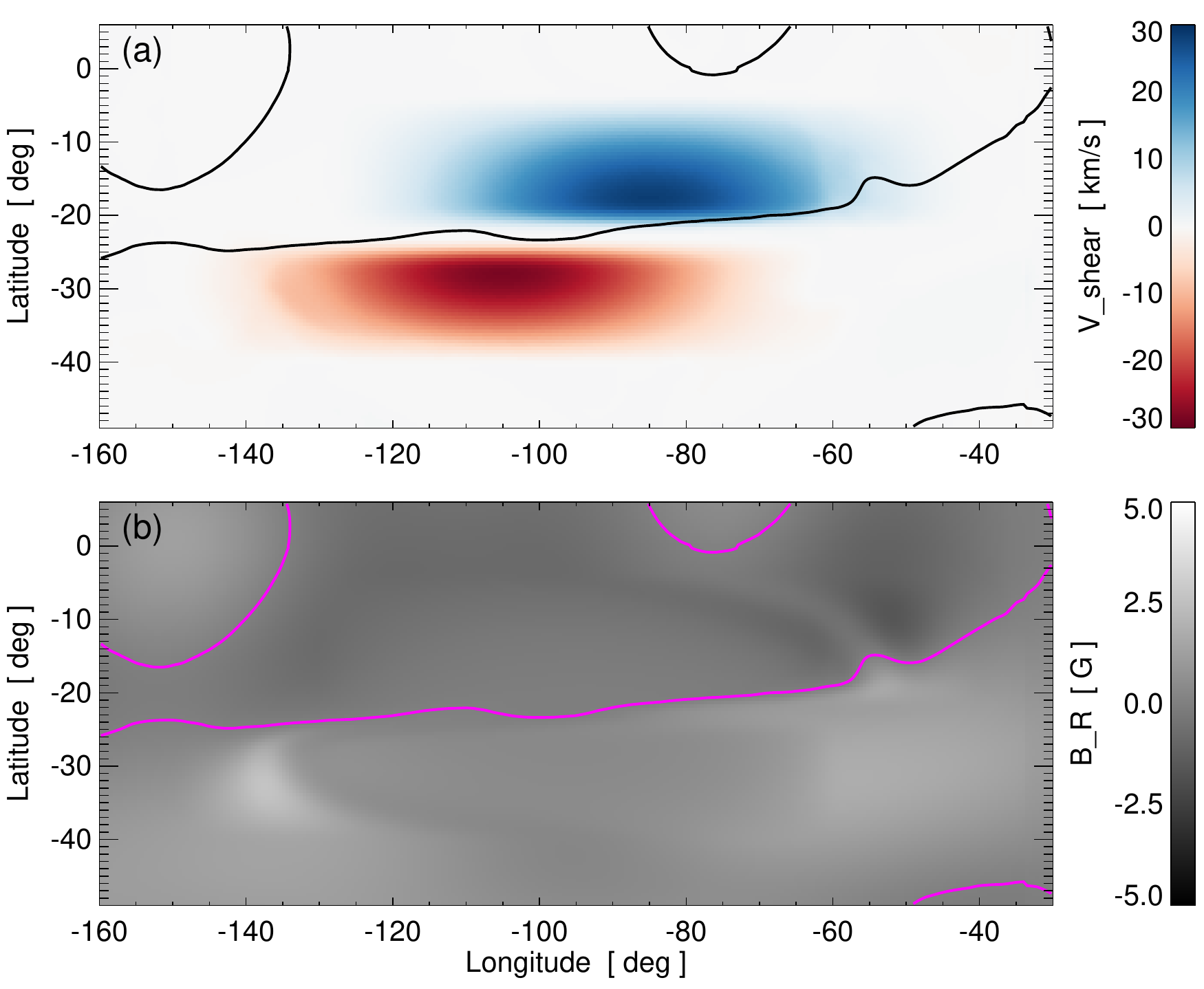}
\caption{(a) The maximum shearing velocity $\boldsymbol{V}_{\rm
	 shear}(t=70~{\rm h})$ on the $r=R_\odot$ boundary. (b)
	 The corresponding $B_r(R_\odot,\theta,\phi)$ distribution
	 at $t=70$~h.  The polarity inversion lines ($B_r=0$
	 contours) are shown as black (Figure~\ref{f6}a) and magenta (Figure~\ref{f6}b) lines.
         \label{f6}
}
\end{figure}

Figure~\ref{f6} plots the spatial extent of the shearing flow profile
imposed at the $r=R_\odot$ lower boundary.  Figure~\ref{f6} (top) shows
$\boldsymbol{V}_{\rm shear}(\theta,\phi,t=70{\rm \; h})$ corresponding
to the time of the maximum shearing velocities.  Figure~\ref{f6} (bottom) 
plots the corresponding $B_r(\theta,\phi,t)$ distribution at the
lower boundary.  The $B_r=0$ magnetic PILs are shown in Figure~\ref{f6} (top)
(Figure~\ref{f6}, bottom) panel as black (magenta) contours. The imposed boundary
flows alter the $B_r$ distribution, concentrating the negative and
positive polarity radial flux, but the field strengths remain low
over the region, i.e., $B_r \in [-5,+5]$~G. Because
the shearing flows alter the PIL, there is a small amount of flux
cancelation during the shearing phase and afterwards due to numerical
diffusion. We have calculated the radial flux in the region depicted
in Figure~\ref{f6} and find the magnitudes of the positive and
negative radial fluxes decrease by less than 2\% during the shearing
phase ($60 \le t \le 80$~h) and less than 0.5\% for $t > 100$~h.
We also note that the \citet{Linker2003PhPl} simulations of flux
cancelation CME initiation typically required a $>10$\% decrease
in the radial flux in order to erupt.

Our applied boundary flows are large scale in order to capture the
effect of shearing by differential rotation, and consequently, they
do not result in the formation of a filament channel.  This is a
key difference between our stealth CME (and perhaps all streamer
blowout CMEs) and the standard fast CME events originating in active
regions that are considered by most CME models
\citep[e.g.,][]{Forbes2006SSR}.  In one class of CME models the
filament channel is presumed to be a twisted flux rope that becomes
ideally unstable. As will be evident below, there is no structure
in our preeruption corona that resembles a flux rope; the whole
closed field system is sheared and there is no significant
flux cancelation.  This makes it highly unlikely that ideal
mechanisms such as the kink or torus instability play a role in the
ensuing eruption.

The simulation presented here also differs from the
\citet{Lynch2009} simulation of 3-D magnetic breakout CME initiation.
The source region field configuration in \citet{Lynch2009} had
higher field strengths ($\pm25$~G) and the shearing flows were more
compact with respect to the active region (AR) PIL. The main topological difference
is in the structure of the background field. In \citet{Lynch2009},
the background fields were a global-scale, closed field dipole above
an oppositely oriented AR bipole resulting in a source region with
a true multipolar topology defined by a 3-D coronal null point and
a separatrix dome surface between the two closed flux systems. Here,
our 3-D helmet streamer is a traditional bipolar arcade flux system
with a topological Y point (line) at the interface between the
closed streamer flux and open flux from polar coronal holes.  The
\citet{Lynch2009} configuration was an idealized representation of
a complex, multipolar AR source, whereas here our stealth CME
originates from a slightly less idealized representation of the
large-scale streamer belt above a quiet-Sun bipolar flux distribution.

Our initial PFSS representation of the large-scale coronal magnetic
field for Carrington Rotation 2070 does not contain any of the
necessary free magnetic energy to power the streamer blowout CME.
One could, in principle, start with an earlier Carrington Rotation
synoptic map and apply flux transport models \citep[e.g.][]{Wang1989,
Schrijver2001} to the entire surface. Variations of this procedure
have been used to successfully drive magnetofrictional modeling of
coronal evolution over a wide range of spatial and temporal scales
\citep[see][]{vanBallegooijen1998, Yeates2009, Cheung2012}.
However, it would be wholly impractical to simulate
many weeks of actual, slow photospheric motions with an explicit
MHD model. Therefore, we are not attempting to mimic the exact
photospheric evolution leading up to this particular CME.  In
general, arbitrarily shaped boundary shear flows can be prescribed
to introduce free magnetic energy into an MHD system
\citep[e.g.,][]{Roussev2007}, including flows that preserve the
radial flux \citep{DeVore2008}.  Here our shearing profiles are
chosen to energize the source region streamer belt while being
constrained and inspired by the relative scales and displacement
associated with 2 weeks of differential rotation.

One consequence of choosing our boundary shearing profiles to be
in the direction of differential rotation is that we generate a
left-handed shear in the helmet streamer arcade, whereas the 6--7 June 2008
ICME flux rope associated with the 1--2 June 2008 CME was
observed to have a right-handed in situ chirality \citep{Moestl2009,
Lynch2010, Nieves-Chinchilla2011}.  As discussed
by \citet{DeVore2000}, a bipolar active region that emerges with a
north-south PIL (the sunspot pair has the same latitude, separated
in longitude) experiencing differential rotation produces a sheared
arcade with the correct hemispheric chirality (left handed for the
Northern Hemisphere, right handed for the Southern Hemisphere). As
the AR decays and the differential shearing continues, the PIL
becomes increasingly titled and oriented along the east-west
direction.  If a bipolar active region emerges with an east-west
oriented PIL (i.e., the sunspot pair emerge at the same longitude,
separated only in latitude), then the same differential rotation
profile produces a sheared arcade with a handedness opposite that
of the hemispheric chirality trend.  In both cases the final state
is an elongated east-west oriented PIL between relatively diffuse
radial field concentrations---the only major difference is the
handedness of the sheared arcade.

The principal limitation of our modeling is that
the CR2070 PFSS configuration used to initialize the MHD simulation
already contains the large-scale PIL oriented in the east-west
direction but none of the previously accumulated shear (of the
correct handedness) that the real Sun must have generated beforehand.
There are always trade-offs in attempting to model the physical
processes and evolution of the real Sun. Since the handedness problem
is well known, in that many authors have discussed this observationally
and in simulations \citep[e.g.][]{DeVore2000, vanBallegooijen1990,
Rust1994, Rust1997}, we chose to generate the wrong handedness over
having to prescribe the boundary flows in the wrong direction (i.e.,
opposite to that of differential rotation).  The handedness
disagreement between our simulation and the observed ICME chirality
implies that this particular region of bipolar quiet-Sun streamer
belt flux and its associated large-scale PIL likely originate from
one or more long decayed active region flux systems.

\subsection{A Slow Streamer Blowout ``Stealth Eruption''}
\label{ssxn.obs.cme}

\subsubsection{Global Evolution and CME Initiation}

\begin{figure}
\includegraphics[width=6.5in]{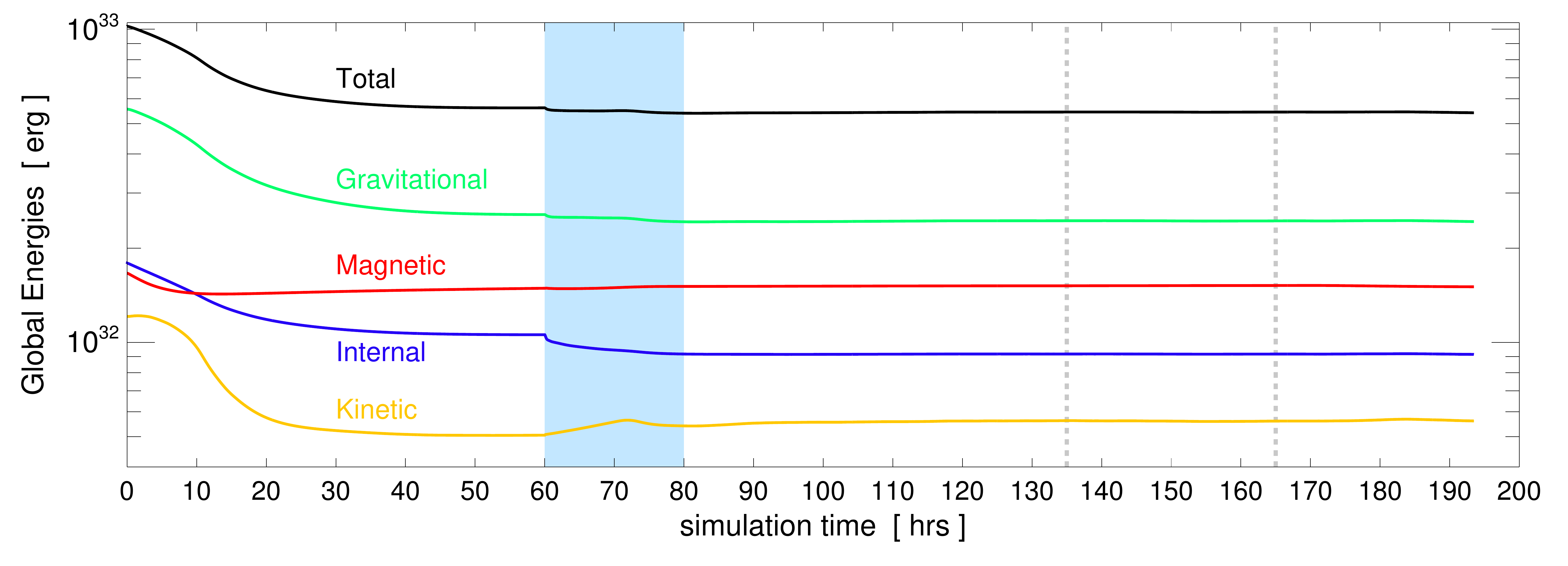}
\caption{Global energy evolution: total (black), gravitational
	 (green), magnetic (red), internal (blue), and kinetic
	 (yellow). The solar wind relaxation phase duration is from
	 $0 \le t \le 60$~h.  The shearing phase is $60 \le t \le
	 80$~h (shaded light blue). The vertical dashed lines
	 indicate the start of the synthetic running difference
	 movie (Figure~\ref{f8}) at $t=135$~h and when the leading
	 edge of the CME enters the COR2A field of view ($r=2.5$R$_\odot$)
	 at $t=165$~h.
         \label{f7}
}
\end{figure}

Figure~\ref{f7} plots the full energy evolution of the MHD system
over the 194~h of simulation time.  We have plotted the total
(black), gravitational (green), magnetic (red), internal (blue),
and kinetic (orange) energies. For $t \le 60$~h the system relaxes
until the isothermal solar wind solution and the initial PFSS
magnetic configuration equilibrate. The values at the end of the
relaxation phase are
$E_{\rm tot} = 5.611 \times 10^{32}$~ergs, 
$E_{\rm grv} = 2.559 \times 10^{32}$~ergs,
$E_{\rm mag} = 1.489 \times 10^{32}$~ergs,
$E_{\rm int} = 1.058 \times 10^{32}$~ergs, and
$E_{\rm kin} = 5.054 \times 10^{31}$~ergs.

\begin{figure}
\includegraphics[width=6.5in]{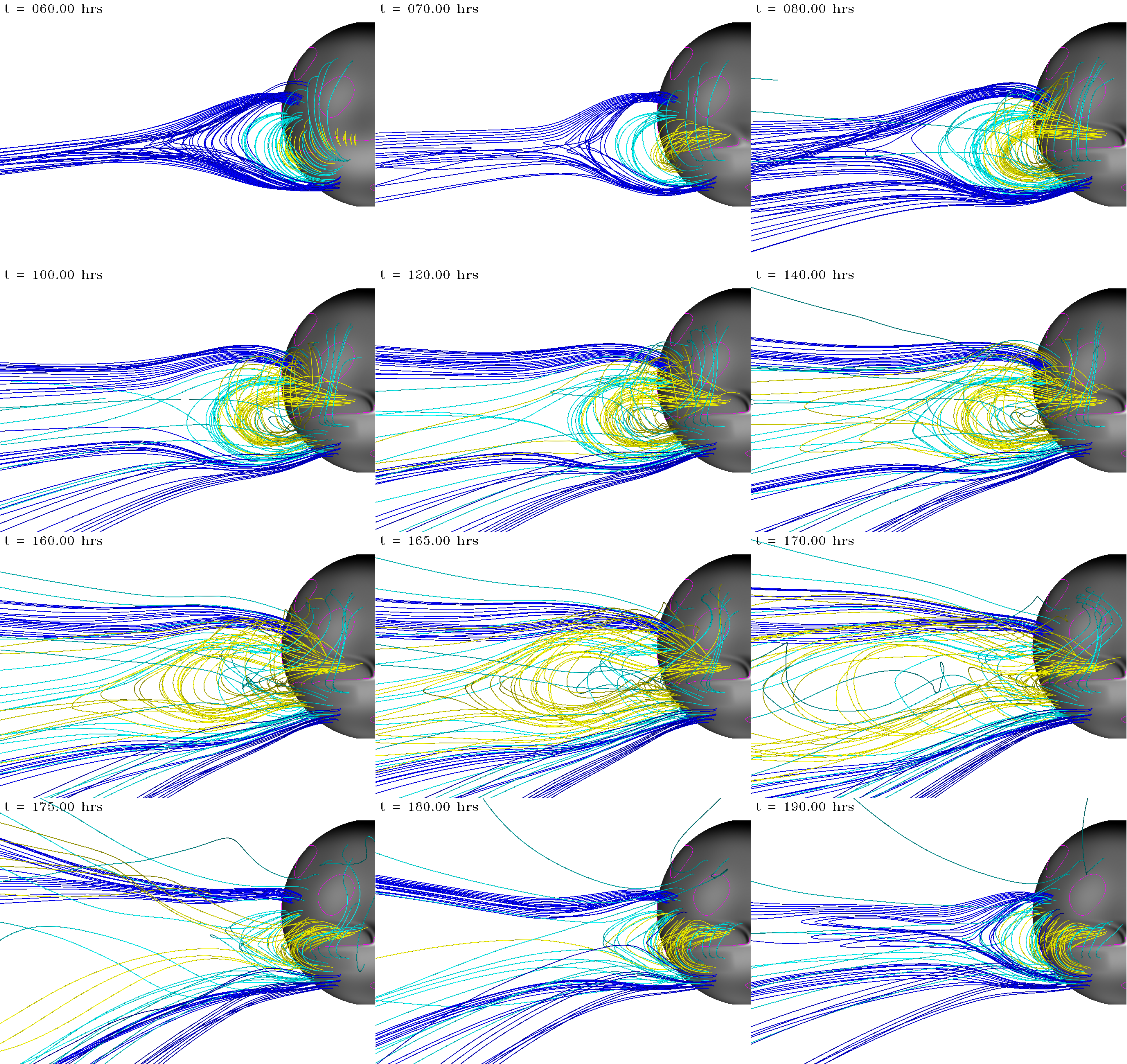}
\caption{Visualization of the magnetic field evolution. The magnetic
	 field lines are colored by their connectivity at $t=60$~h:
	 Yellow field lines are traced from points within the shear
	 channel and become the highly sheared core flux that
	 eventually erupts; light blue field lines show the restraining
	 closed flux of the helmet streamer outside of the shear
	 channel; dark blue field lines show the boundary of the
	 helmet streamer and adjacent open fields. For $t \ge 80$~h,
	 all of the field lines are traced from static foot points
	 showing the topological evolution of the stealth streamer
	 blowout CME. An animation of this figure is available as
	 supporting information of the article.
         \label{f10}
}
\end{figure}

Figure~\ref{f10} shows representative magnetic field lines during
the simulation. The magnetic field lines are colored by their
magnetic topology at $t=60$~h.  The dark blue field lines show the
largest-scale streamer field lines and open fields just outside the
helmet streamer boundary.  The light blue field lines show the
overlying closed streamer flux that originates outside of the shear
channel.  The yellow field lines are traced from positions within
the shear channel and show the energized arcade flux.

The top row in Figure~\ref{f10} shows the shearing of the bipolar
arcade for $t \in [60, 80]$~h. The magnetic flux distribution on
the lower radial boundary remains fixed for $t \ge 80$~h.
The second row shows the slow arcade expansion
during $100 {\; \rm h} \le t \le 140 {\; \rm h}$. As the arcade
expands, the light blue flux gradually opens into
the solar wind. As this overlying flux opens up, the restraining
tension force decreases, allowing the sheared flux core to expand
further in a positive-feedback loop reminiscent of the magnetic
breakout \citep{Antiochos1999, Lynch2008} or tether-cutting
\citep{Moore2001} CME initiation mechanisms. Rather than reconnecting
at an overlying breakout current sheet and being transferred to
adjacent arcades as in the classic breakout scenario, here the
restraining flux gradually expands into solar wind to become open.
The sheared arcade expansion forms a radial current sheet  that
elongates, thins and eventually facilitates magnetic reconnection
in the standard CSHKP scenario.  The third row in
Figure~\ref{f10} shows the onset of eruptive flare reconnection for
$t \gtrsim 160$~h. This magnetic field reconfiguration creates a
flux rope-like structure \emph{during} the eruption process.  The
bottom row in Figure~\ref{f10} shows the posteruption reconnection
that reforms the helmet streamer beneath the CME for $t \gtrsim
175$~h.

Figure~\ref{f7zoom} (top) plots the
global magnetic and kinetic energies during the eruption normalized
to their respective values at $t=170$~h. The shape of the energy
curves are exactly as expected from our previous simulations of CME
events: the magnetic energy decreases and the kinetic energy increases
as the erupting fields are removed from the low corona. 
We note that for $t \gtrsim 187$~h, some portion of both the global 
magnetic and kinetic energy decrease is due to the CME passing through 
the simulation outer boundary.
Figure~\ref{f7zoom} (bottom) shows the height-time evolution of the
leading edge of the expanding arcade and CME derived from the
analysis discussed in section~\ref{ssxn.ht}.  The height-time profile
shows the smooth transition for $165 \lesssim t \lesssim 175$~h
from the linear rising phase of the sheared arcade to the full CME
eruption and its associated increase in the global kinetic energy.
This transition corresponds to the onset and development of eruptive
flare reconnection and its rapid reconfiguration of the magnetic
field connectivity (as seen in the $t=160$, $165$, and $170$~h
panels in Figure~\ref{f10}).  The relative timing of the onset of
flare reconnection and the resulting acceleration during the
transition from expanding arcade to erupting flux rope suggests that our
eruption mechanism is that of a resistive instability in the radial
current sheet.

There are other ideal MHD instabilities that can
initiate CME eruptions such as loss-of-equilibrium \citep{Forbes2006SSR},
in which the system suffers a catastrophic change in the equilibrium
states available to it, and the torus instability \citep{Kliem2006},
in which an equilibrium flux rope becomes ideally unstable to outward
displacements into a region of substantially weaker field. In either
of these ideal instability cases, we would expect to observe the
onset of outward motion \emph{before} flare reconnection sets in
below the rising sheared arcade, whereas we observe the outward
acceleration only \emph{after} the flare reconnection sets in.
Additionally, ideal MHD instabilities require a net
current, in other words, twist, which is either obtained by the
presence of a preeruption twisted flux rope \citep[e.g.,][]{Torok2005,
Kliem2006, Fan2010} or created via flux cancelation
\citep[][]{Linker2003PhPl, Amari2010, Aulanier2014IAU}.  There has
never been an ``ideal eruption'' of a sheared arcade, nor can there
be, irrespective of how high the sheared flux expands to.  The key
point is that the flare reconnection allows the system to evolve
to a much lower energy state than it could ever get to by ideal
expansion.

Therefore, despite the slow, gradual character of
our stealth CME eruption, the main acceleration mechanism is the
same as in the fast 2.5D magnetic breakout CME discussed by
\citet{Karpen2012}: the onset of flare reconnection forms and
accelerates the CME flux rope.  The key property of the evolution
that is responsible for eruption is that the outward expansion
results in the formation of a large, vertical current sheet deep
inside the closed field system. As concluded by \citet{Karpen2012},
once reconnection sets in at this flare current sheet, the system
must erupt with the ejection of a large flux rope.  This conclusion
appears to apply to our stealth CME as well, even though the system
is fully 3-D.  Without reconnection, the sheared flux could open
(``erupt'') via ideal expansion but the large vertical current sheet
would extend down very close to the surface. This open state has
much greater magnetic energy than the potential state. The eruptive
flare reconnection allows the system to take the difference between
the energy of this open state and the potential state and put it
into accelerating the CME plasma and opening up a small fraction
of the low-lying flux.  It is unclear whether smooth flux opening
by ideal expansion would be classified as a CME by an observer.

The flare reconnection is essential to the CME
eruption in that the CME's magnetic structure is a consequence of
this reconnection. Specifically, (1) the eruptive flare reconnection
creates a flux rope structure from the expanding arcade, and (2)
this reconnection enables the removal of a significant amount of
sheared (and due to reconnection, increasingly twisted) flux from
the closed field corona as a coherent structure.  Without the flare
reconnection, the opening flux will be substantially less ``coherently
structured'' and will not exhibit the flux rope-like cross-sectional
morphology in coronagraph images. At the same time,
we show that for this stealth CME, the acceleration during an
eruption with relatively weak flare reconnection can be small.
There is some acceleration in our simulation, the global kinetic
energy does increase from 170--184~h, but it is not very substantial.
Consequently, the resulting CME speeds are comparable to the
background solar wind.

There is a hierarchy of energetics associated with
reconnection-driven eruption of closed flux. Large flares that
generate fast CMEs originate in very low lying fields above polarity
inversion lines of active regions. Streamer blobs are generated at
the edges of the closed flux regions, near the open-closed field
interface, i.e. relatively weak fields at the boundaries of coronal
holes and/or even weaker fields in the extended corona. Stealth
CMEs, and likely all slow streamer blowout CMEs, are essentially
an intermediate case between these two extremes. They originate at
intermediate heights in the closed field regions and thus have a
much larger spatial extent than streamer blobs but are much less
energetic than fast CMEs.  Magnetic reconnection is required for
all of these structures to ``erupt'' (be released) into the open
field \emph{with a coherent flux rope-like magnetic structure}.

\subsubsection{Eruptive Flare--CME Energy Budget}

\begin{figure}
\center
\includegraphics[width=6.5in]{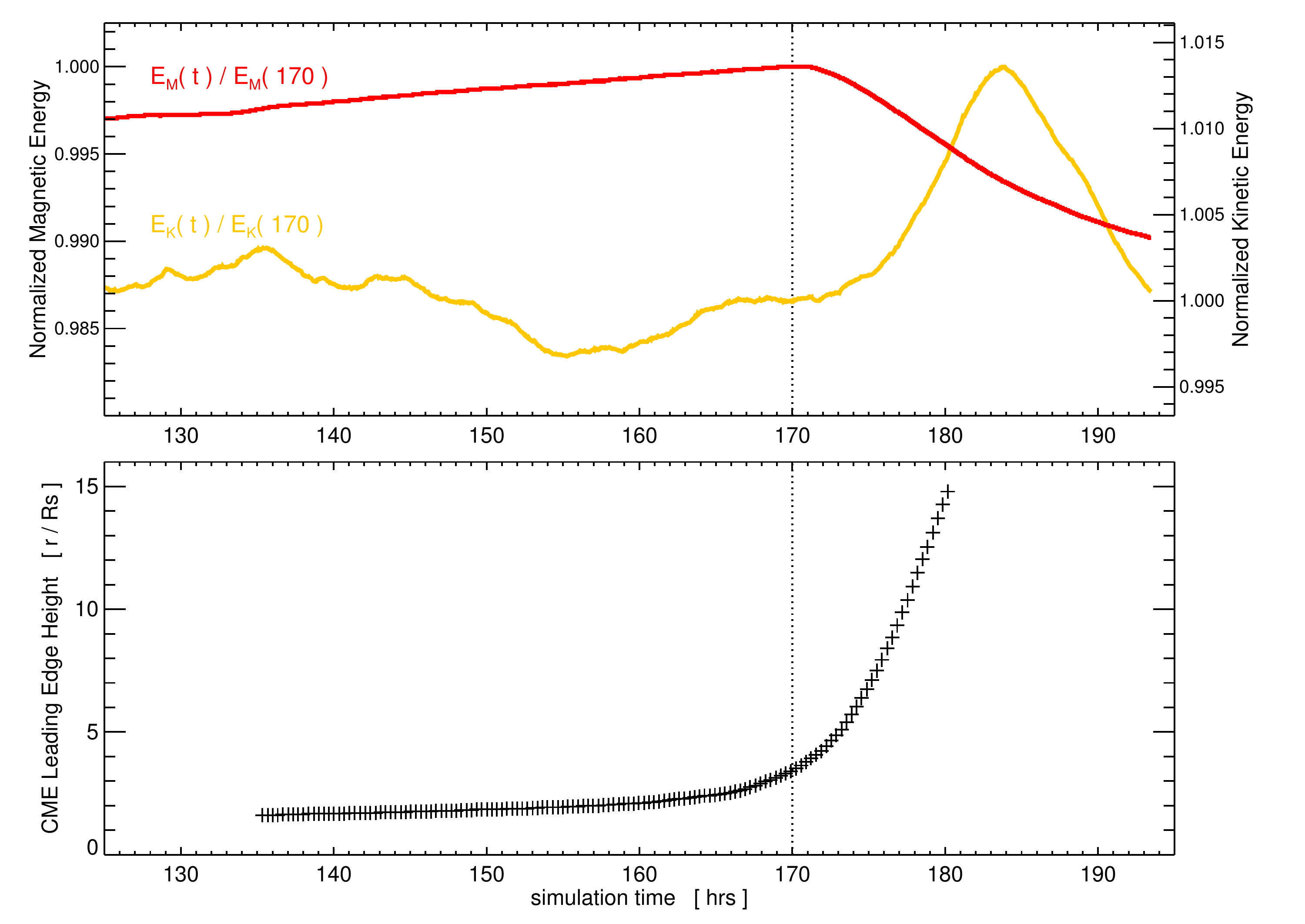}
\caption{(top) Normalized global magnetic
	 and kinetic energies during the slow stealth CME eruption.
	 We normalize by the values at $t=170$~h (dashed vertical
	 line) where the eruptive flare reconnection is most visible
	 in Figure~\ref{f10}.  The stealth CME eruption corresponds
	 to only a 1.0\% drop in total magnetic energy
	 ($1.49\times10^{30}$~ergs) and a 1.4\% increase in total
	 kinetic energy ($7.62\times10^{29}$~ergs). (bottom) Height-time evolution of the sheared arcade expansion
	 and slow CME eruption.
         \label{f7zoom}
}
\end{figure}

Our CME simulation can be considered a ``stealth''
event because the magnitude of both the magnetic energy decrease
and the CME's kinetic energy increase are relatively small. From
Figure~\ref{f7zoom}, the global magnetic energy released over the
course of the eruption is $\sim$1.0\% ($\Delta E_M =
1.49\times10^{30}$~ergs) and the peak of the global kinetic energy
is $\sim$1.4\% ($\Delta E_K = 7.62\times10^{29}$~ergs) compared to
the preeruption values.  The temporal and spatial scales over which
the released magnetic energy is deposited are also extremely important
in characterizing the eruptive flare/CME event dynamics.

\begin{figure}
\center
\includegraphics[width=6.5in]{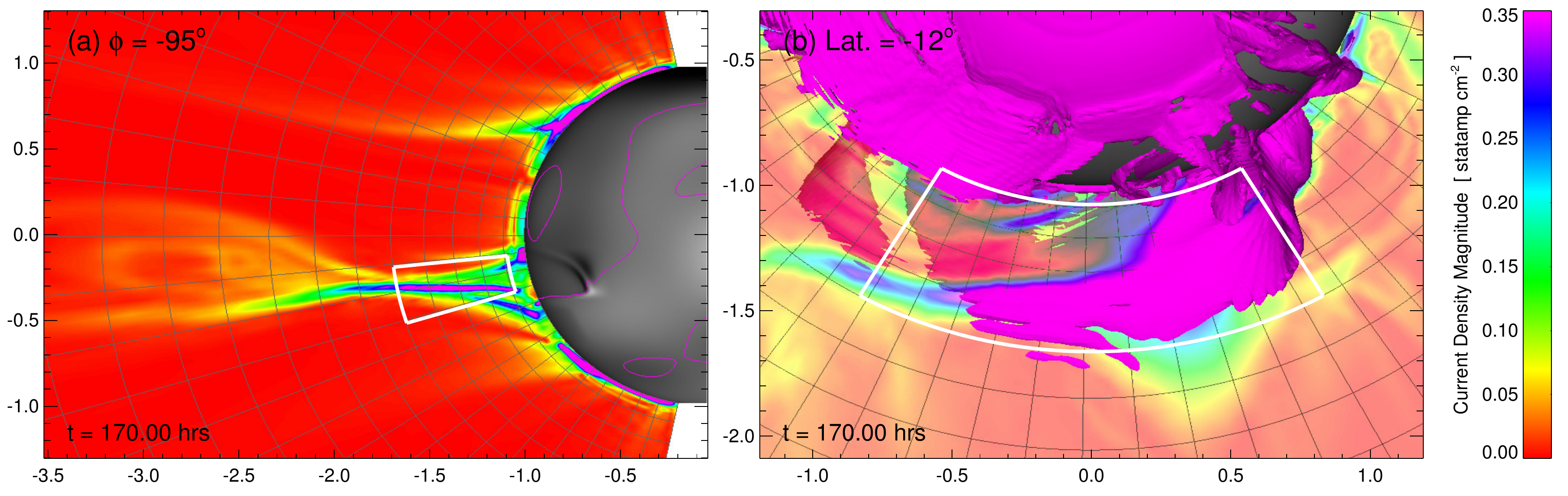}
\caption{Current density magnitude $|J|$ of the eruptive flare
	 current sheet at $t=170$~h. (a) The $r$-$\theta$ plane at
	 $\phi=-95^\circ$.  (b) Transparent $r$-$\phi$ plane at
	 latitude $-12^\circ$ with the magenta isosurface at
	 $|J|=0.35$~statamp~cm$^{-2}$. The white spherical wedge
	 regions indicate the radial and azimuthal areas used in
	 the equation~(\ref{eqS}) estimate of the energy density flux
	 available to generate plasma heating and flare emission
	 signatures.
         \label{fcs}
}
\end{figure}

An upper limit estimate of the Poynting flux, $\tilde{S}$, into the
stealth CME flare current sheet can be obtained via
\begin{equation}
    \frac{\partial E_M}{\partial t} = \oint \boldsymbol{S} \cdot
    {d\boldsymbol{A}} \ \Longrightarrow\ \frac{\Delta E_M}{\Delta
    t} \sim \tilde{S} \Delta A .
\end{equation}
Figure~\ref{fcs} shows two viewpoints of the current
density magnitude $|J|$ in the flare current sheet at $t=170$~h:
(a) in the $r$-$\theta$ plane at $\phi=-95^\circ$; and (b) in the
(transparent) $r$-$\phi$ plane at latitude $-12^\circ$.
Figure~\ref{fcs}b also plots the 3-D isosurface of $|J|=0.35$~statamp~cm$^{-2}$.
An estimate of the $r$-$\phi$ planar wedge area associated with the
stealth CME current sheet is calculated from $r \in
[1.1R_\odot,1.7R_\odot]$ and $\phi \in [-65^\circ,-125^\circ]$ at
latitude $\lambda = -12^\circ$. This gives an area of $\Delta A =
4.21 \times 10^{21}$~cm$^2$.  Taking the duration of the magnetic
energy drop as 24~h from Figure~\ref{f7zoom} ($\Delta t = 8.64
\times 10^4$~s), we get
\begin{equation}
    \tilde{S} = \frac{(1.49 \times 10^{30} {\; \rm erg})}{(4.21
    \times 10^{21}{\; \rm cm}^2)(8.64 \times 10^{4}{\; \rm s})} =
    4.10 \times 10^{3}{\; \rm erg \; cm^{-2} \; s^{-1}}.
    \label{eqS}
\end{equation}
This estimate of the energy flux into the stealth CME's
large-scale current sheet over the course of the eruption is about
2 orders of magnitude below the average energy flux necessary to
heat the ambient solar corona and accelerate the solar wind
($\sim$$10^{5-6}$~erg~cm$^{-2}$~s$^{-1}$) \citep[e.g.,][]{Withbroe1977,
Fisk1999b, Cranmer2010}.  Therefore, it is not at all surprising
that the 1--2 June 2008 stealth CME event did not produce any
significant, or observable, flare emission and posteruption
brightening (heating) in the low corona.

\subsection{Comparison to Observations}
\label{ssxn.obs.fluxrope}

\subsubsection{Current Sheet X Point Structure}

\begin{figure}
\center
\includegraphics[width=4.25in]{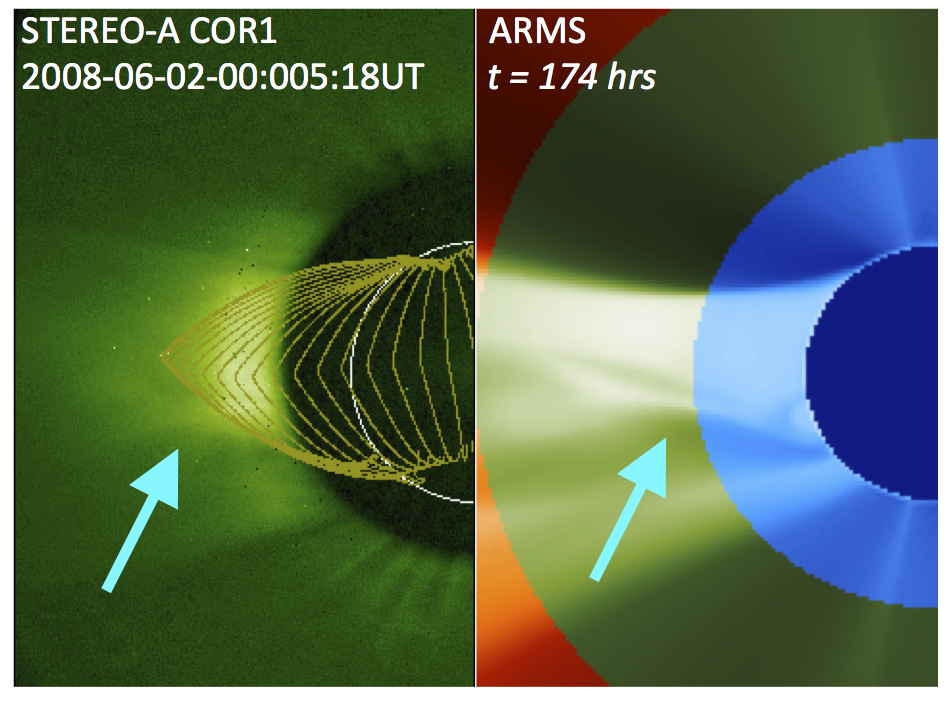}
\caption{(left) STA-COR1 observation with representative PFSS
	 field lines overplotted showing the faint white-light
	 signature of the eruption X point \citep[adapted
	 from][]{Lynch2010}. (right) Synthetic white-light
	 ratio image at $t=174$~h from $\phi_{\rm STA}$ perspective
	 showing the stealth CME simulation X point structure.
         \label{fxpt}
}
\end{figure}

Current sheets trailing CME eruptions have been observed in white-light 
coronagraph observations \citep[e.g.,][]{Webb2003, Lin2005ApJ,
Poletto2008, Patsourakos2011AA, Song2012SoPh}  as well as EUV imaging
and spectroscopy and X-rays \citep[e.g.,][]{Bemporad2006ApJ,
Savage2010, Liu2010ApJ, Liu2013MNRAS, Landi2012ApJ, Ciaravella2013,
Susino2013ApJ}.  Given the upper limit estimate of the magnetic
energy flux available for bulk plasma heating and particle acceleration
associated with our stealth CME current sheet, we do not expect any
of the typical coronal EUV or X-ray emission signatures associated
with high-energy eruptive flare events.  
Figure~\ref{fxpt} compares
the faint white-light observation of the CME-trailing eruptive flare
current sheet and X point morphology in the STEREO-A COR1 coronagraph
(Figure~\ref{fxpt}, left) with the corresponding synthetic white-light emission
calculated from the MHD simulation (Figure~\ref{fxpt}, right). 
The intensity enhancement
suggestive of the X point morphology is quite high in the corona,
$\sim$1.5$R_\odot$. There is good correspondence between the
white-light intensity structures observed by COR1A and that generated
from the simulation data.

\subsubsection{Three-Part CME Structure}

The three-part morphology of CMEs in white-light coronagraph
observations is one of the best proxy measures of the plasma and
magnetic field structure of CMEs \citep[e.g.,][]{Illing1985, Dere1999,
Wood1999, Cremades2004, Rouillard2011JASTP, Vourlidas2013}.  Here
we present a favorable comparison between the observed large-scale
white-light morphology of the 1--2 June 2008 CME and the synthetic
white-light structure derived from the simulated flux rope CME
density distribution.

\begin{figure}
\includegraphics[width=6.5in]{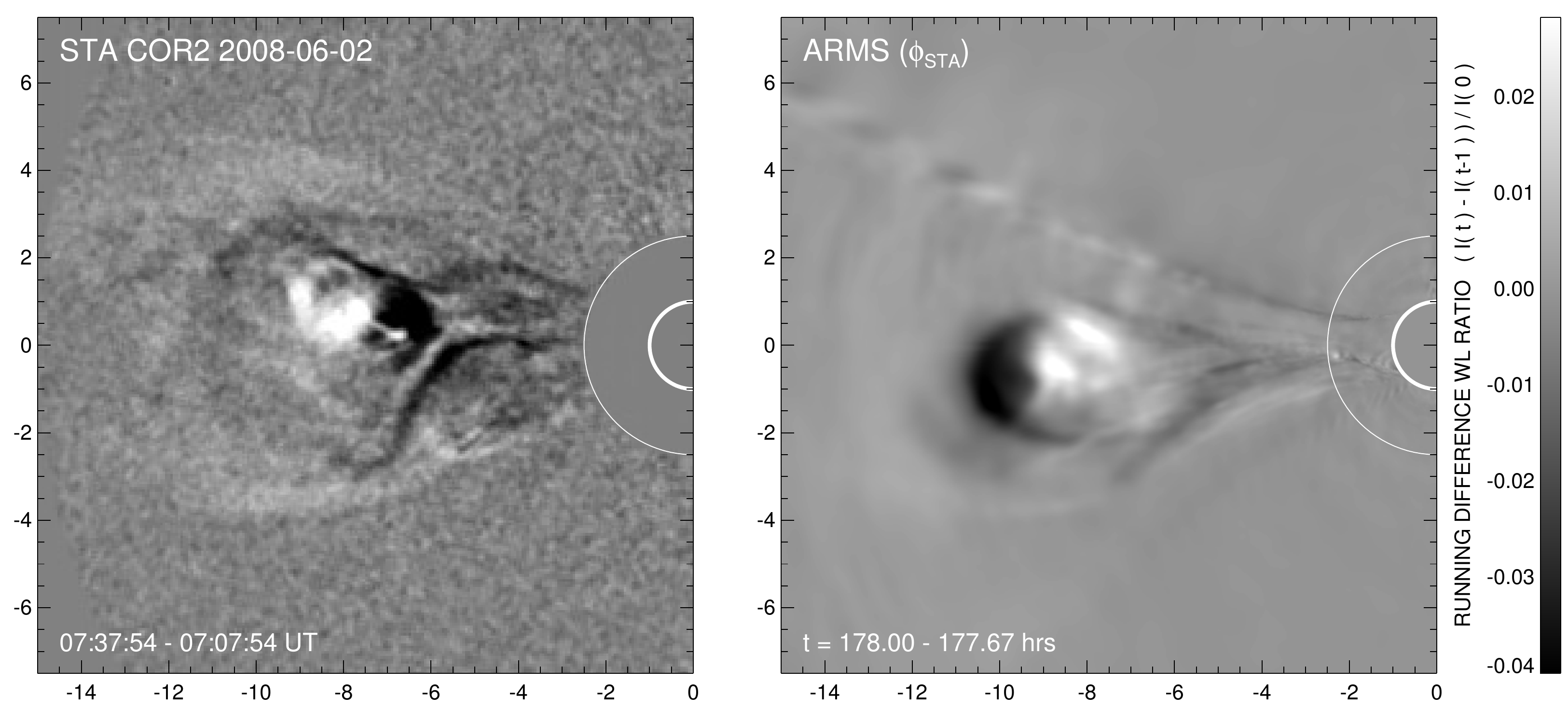}
\caption{(left) STA-COR2 running difference image at 07:37UT
	 on 2 June 2008 showing the three-part white-light structure of
	 the flux rope CME. (right) Running difference image
	 of synthetic white-light ratio images at $t=178$~h showing
	 the three-part structure in the simulated CME. An animation
	 of the simulated running difference evolution is available
	 as supporting information of the article.
         \label{f8}
}
\end{figure}

Figure~\ref{f8} plots the STA COR2 running difference image showing
the circular cross section of the stealth CME at $\sim$10$R_\odot$
(Figure~\ref{f8}, left) and the corresponding simulation results (Figure~\ref{f8}, right).
The simulation running difference images are
calculated from the synthetic total white-light brightness ratio
images as $( I( t ) - I( t-1) ) / I(0)$ using simulation output
files generated every 20~min of simulation time. The opening angle
of the eruptive structure and the relative size of the circular
cross section are in excellent qualitative agreement.  Our simulation
also produces the bright core density enhancement at the center of
the flux rope. This is an improvement over our earlier 2.5-D
\citep{Lynch2004} and 3-D \citep{Vourlidas2013} results. The solar
wind boundary conditions used here allow additional mass to accumulate
in the closed field region before being transferred by the flare
reconnection outflow into the erupting flux rope structure.  On the
other hand, our simulation shows the running difference signature
of a dark cavity ahead of the bright core, whereas in the STA COR2
observations the running difference signature of the cavity region
is much less pronounced.  This seems to be due to the simulated
flux rope cavity being more depleted relative to the ambient streamer
density than the observed streamer blowout CME. In fact, many CMEs
generally do show the well-known three-part structure of a bright
front followed by a dark cavity and then a bright core, but the
structure appears to be reversed in this stealth event. This is
likely due to the long buildup phase of this CME, which allowed for
substantial mass transfer from the corona into the ejecting plasmoid
during its development. Our shearing phase, however, is approximately
20 times faster than the Sun's equivalent differential rotation
energization timescale of $\sim$15~days. Furthermore, our use of
an isothermal plasma energy equation is likely to underestimate the
effects of plasma heating and chromospheric evaporation.

\begin{figure}
\includegraphics[width=6.0in]{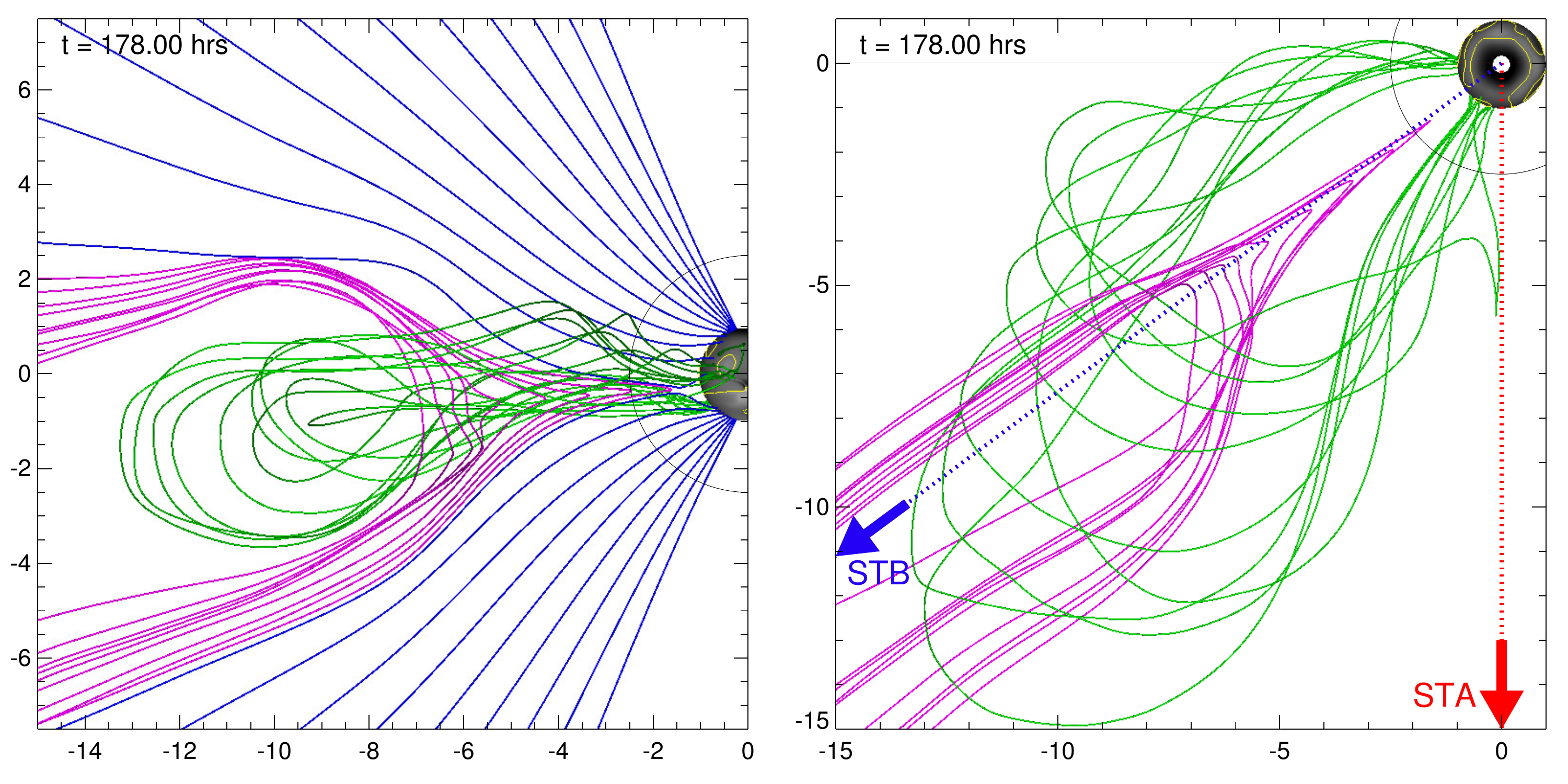}
\caption{(left) Representative magnetic field lines of the
	 stealth CME eruption for the ARMS $\phi_{\rm STA}$
	 field of view in Figure~\ref{f8} . Shown are the sheared
	 arcade field lines that become the flux rope CME (green),
	 the reconnecting open fields behind the eruption (magenta),
	 and the background open field lines (blue). (right)
	 The same green and magenta field lines from the North solar
	 pole. The longitudinal positions of the STA and STB
	 spacecraft in the simulation are shown with the red and
	 blue arrows, the STA COR2 plane of the sky as the solid
	 red line.
         \label{f8_t178}
}
\end{figure}

Following \citet{Vourlidas2013}, here we analyze the simulation's
magnetic flux rope structure and topology from the synthetic STA
viewpoint in Figure~\ref{f8}.  Figure~\ref{f8_t178} (left)
plots representative magnetic field lines from the simulation at
$t=178$~h colored by their magnetic connectivity. The erupting
flux rope field lines are shown in green, the newly reconnected
field lines behind the eruption are shown in magenta, and the
background open field lines are shown in blue.  The faint,
running difference leading edge enhancement occurs just ahead of
the approximately circular cross section of the green flux rope
field lines. The green field line cross section corresponds to the
dark running difference cavity. The running difference bright core
enhancement is located at the back of the flux rope at the interface
of the green and magenta field lines, suggesting that it originates
in the large-scale flare reconnection outflow. In fact, the
running-difference Y-shaped feature behind the eruption is now
easily related to the reconnecting fields at the eruptive flare
current sheet.

Figure~\ref{f8_t178} (right) plots the green and magenta CME
lines from the top-down perspective looking at the north solar pole.
The heliospheric longitude position of the STA and STB spacecraft
relative to the simulation data are shown as the red and blue arrows,
respectively. There are two important features.  First, the radial
trajectory of the stealth CME is more directed towards STB rather
than lying in the STA plane of the sky.  Second, the relatively
compact circular cross section of green flux rope field lines seen
in Figure~\ref{f8_t178} (left) is a projection effect---the synthetic COR2A
plane of the sky is shown as the horizontal red line at $y=0$. The
core flux rope field lines have a greater radial extent in the
portion of the flux rope between the STB and STA position angles.
In addition, the 3-D stealth CME structure has a significant large-scale
writhe component in the magnetic field, which can be difficult to
distinguish from twisted flux rope fields in in situ data
\citep[e.g.,][]{Jacobs2009,AlHaddad2011}.  Our
stealth CME magnetic field structure in the coronagraph field of view
is that of a twisted flux rope, just not a \emph{highly
twisted} flux rope. The amount of poloidal twist flux contained in
the erupting magnetic structure should be directly proportional to
the amount of flare reconnection during the eruption, especially
for sheared arcade preeruption states \citep[e.g.,][]{Longcope2007,
Kazachenko2012}. Therefore, it is completely reasonable that our
slow CME having undergone relatively weak flare reconnection ends
up less twisted than more energetic CMEs with greater reconnection
flux during their eruptive flares.

\subsubsection{Height-Time and CME Speed Profile}
\label{ssxn.ht}

\begin{figure}
\includegraphics[width=6.5in]{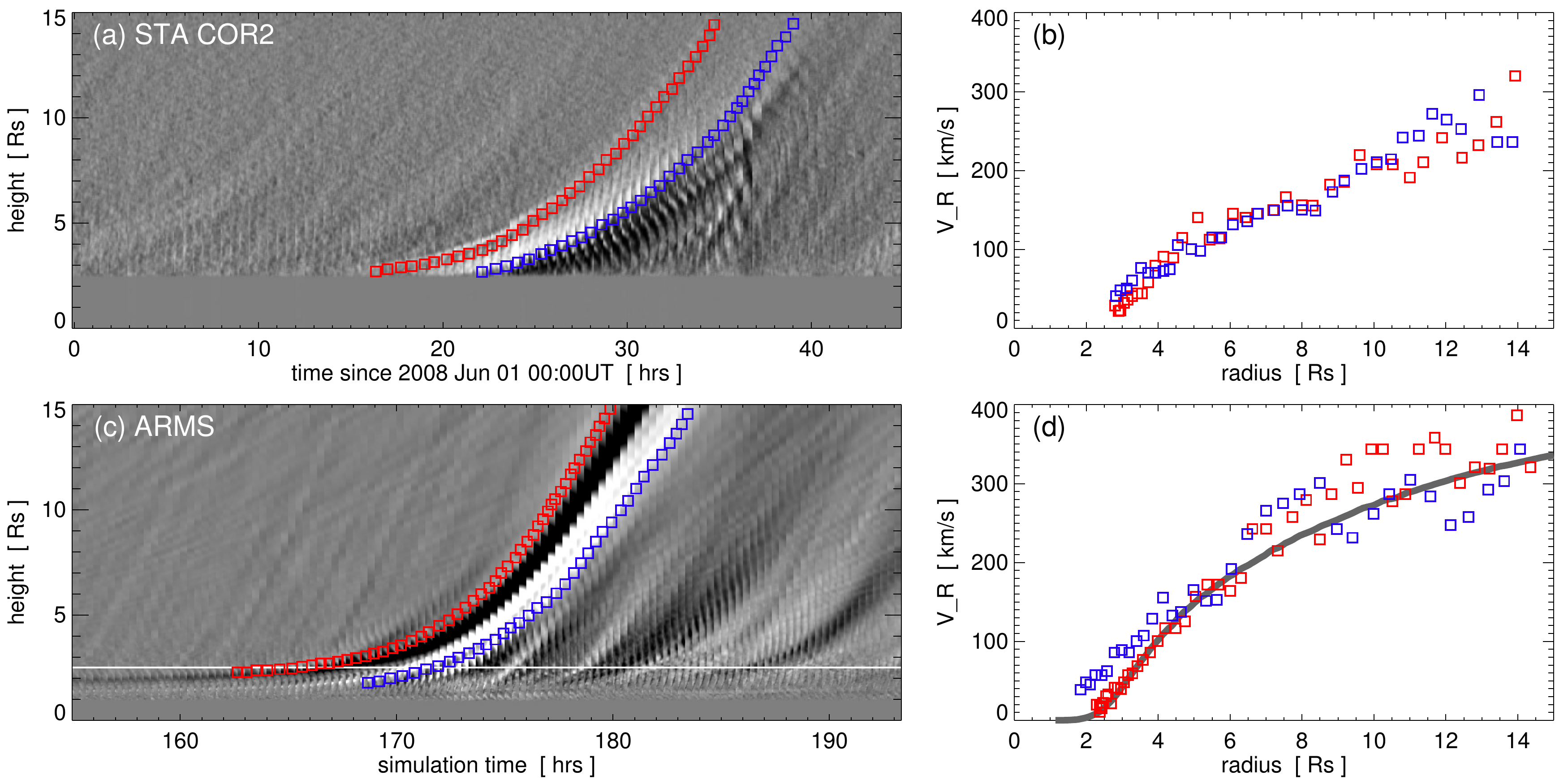}
\caption{(a) STA-COR2 running difference ``J-map'' with height-time
	 points for the CME leading edge (red
	 squares) and core region (blue squares)
	 overplotted. (b) The apparent radial velocity calculated
	 from the height-time tracks as a function of $r$. (c) The
	 simulated height-time ``J-map.'' (d) The simulated CME
	 velocity profile. The black line is the ambient solar wind
	 profile at the equator from Figure~\ref{f3}.
         \label{f9}
}
\end{figure}

\citet{Ma2010} examined the STEREO COR1 and COR2 signatures of slow
CME events both with and without low coronal signatures during solar
minimum (January--August 2009) and found that approximately one third of their
CMEs could be considered stealth events. These events were consistently
at the lower end of the slow CME velocity profile distribution ($V_r
< 300$~km~s$^{-1}$ by $\sim$15$R_\odot$).  \citet{Vourlidas2000}
calculated the velocity and acceleration profiles, as well as the
kinetic energies, of 11 CMEs with flux rope morphologies in LASCO
observations during the previous solar miniumum (1997). Their five
slowest events ($V_r \le 350$~km~s$^{-1}$ by $\sim$20$R_\odot$)
corresponded to kinetic energy estimates in the range of $3-5 \times
10^{29}$~ergs, in good agreement with our simulation results.
Figure~\ref{f9} compares the height-time and velocity profiles for
the 1--2 June 2008 CME observations (Figures~\ref{f9}a and \ref{f9}b) and our simulation
(Figures~\ref{f9}c and \ref{f9}d).  Figures~\ref{f9}a and \ref{f9}c plot the height-time
``J-maps" \citep{Sheeley1997} constructed from the running-differenced
STA COR2 data and our synthetic white-light images. The red squares
track the cavity leading edge and the blue squares
track the trailing edge of the CME core enhancement of the eruption;
these points are used to calculate the apparent $V_{r}(r)$ profiles
plotted in Figures~\ref{f9}b and \ref{f9}d.  In Figure~\ref{f9}d
we also plot the equatorial background solar wind profile from
Figure~\ref{f3} as the black line.  In both the
observations and simulation, the CME core velocities are slightly
higher than the leading edge velocities for $r \lesssim 4 R_\odot$,
indicative of the additional acceleration the core material experiences
from originating in the flare reconnection jets. For $r \gtrsim 4
R_\odot$, this separation has disappeared, implying that, at least
for $r \gtrsim 4 R_\odot$, our stealth CME flux rope is now advecting
passively with the background flow.

\subsubsection{Synthetic In Situ Time Series at 15$R_\odot$}

\begin{figure}
\includegraphics[width=3.5in]{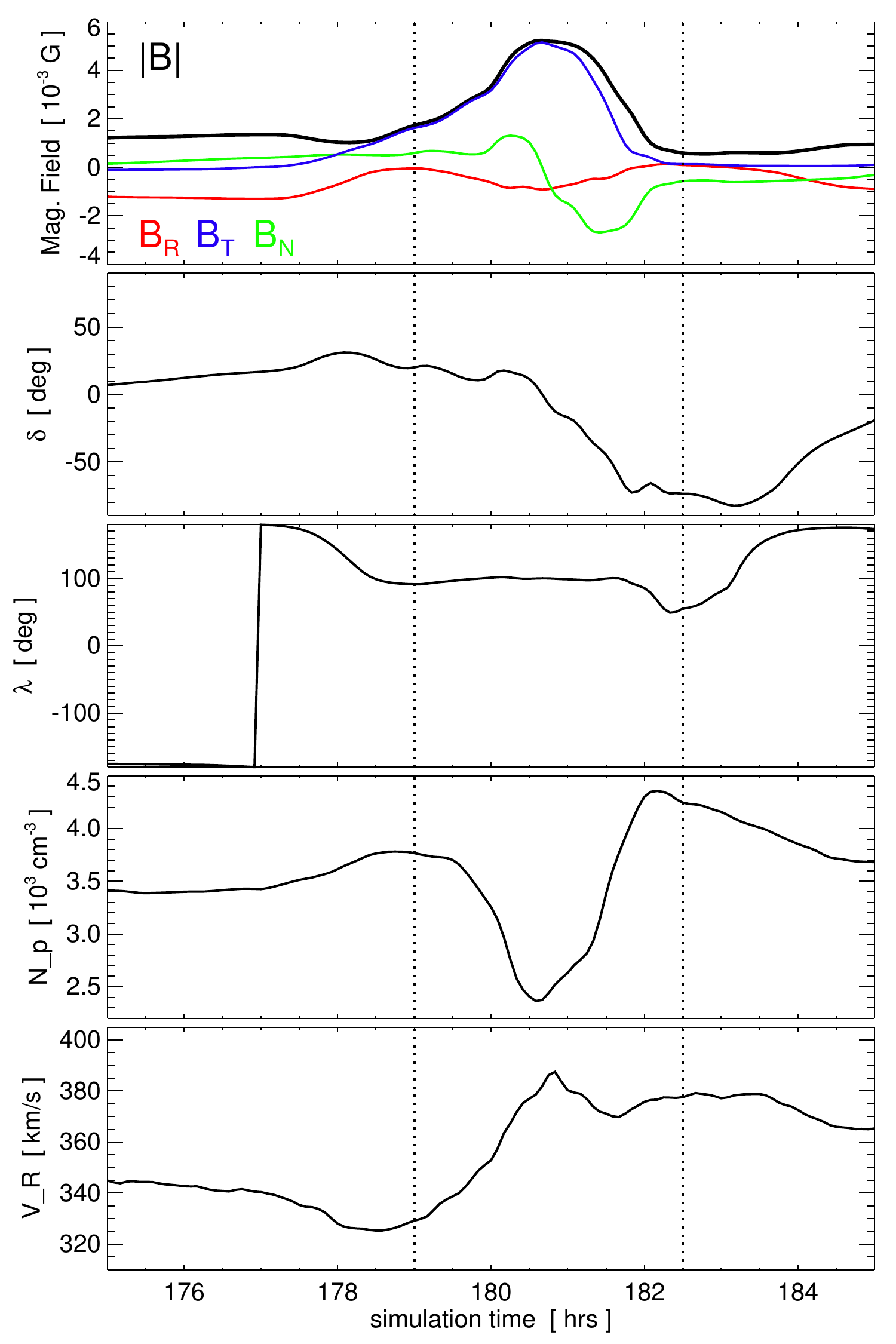}
\caption{Synthetic in situ time series of field and
	 plasma properties at $\boldsymbol{r} = (15R_\odot,\:
	 0^\circ,\: \phi_{\rm STB})$. From top to bottom, $B$
	 magnitude (black) and RTN components ($B_R$ red; $B_T$
	 blue; $B_N$ green), magnetic field direction ($\delta$,
	 $\lambda$), number density $N_p$, and radial velocity
	 $V_r$. Vertical dotted lines denote period of coherent,
	 ICME flux rope rotation.
         \label{f14}
}
\end{figure}

For the time period $t \in \left[ 175, 185\right]$~h
we have increased the temporal cadence of the simulation output
files to once per 5~min to construct the synthetic time series of
plasma and field properties seen by a stationary observer at
$\boldsymbol{r} = (15R_\odot,\: 0^\circ,\: \phi_{\rm STB})$.
Figure~\ref{f14} shows, from top to bottom, magnetic field magnitude
(black) and components in RTN coordinates ($B_R=B_r$, red;
$B_T=B_{\phi}$, blue; $B_N=-B_{\theta}$, green), magnetic field
direction (in RTN latitude $\delta$ and RTN longitude $\lambda$),
plasma number density $N_p$, and radial plasma velocity $V_r$. The
synthetic in situ time series shows many of the classic characteristics
of flux rope ICMEs, often referred to as magnetic clouds, including
the enhanced magnetic field magnitude, a smooth, coherent rotation
in the field direction, and a lower density region corresponding
to a lower in situ temperature \citep{Bothmer1998, Lynch2005, Li2011,
Li2014}.

\section{Discussion}
\label{sxn.disc}

\subsection{Implications for CME Initiation and Large-Scale Coronal Evolution}
\label{ssxn.disc.cme}

Recent observations of coronal mass ejections that lack typical
low coronal signatures, such as EUV or X-ray flaring and ribbons,
have led to them being characterized as a new and mysterious type
of ``stealth CME.''  However, every individual component of our MHD
simulation---the low-order PFSS field representation
\citep[][]{Altshuler1969, Schatten1969, Wang1992}, the isothermal
solar wind model \citep{Parker1958}, and sheared arcade model for
CMEs \citep{Forbes1982, Forbes1983, Mikic1988, Mikic1994, Linker1995,
Antiochos1999}---is 30 to 60 years old. Yet we have shown that
these components, when combined with modern computational techniques,
lead to a 3-D eruption scenario that accounts for a majority of the
qualitative features of the 1--2 June 2008 stealth CME.

The physical processes associated with the gradual shearing
energization and the resulting arcade expansion that lead to
large-scale magnetic reconfiguration with relatively weak (low
energy) eruptive flare reconnection are about as fundamental and
generic as possible---and have been observed for decades. Therefore,
not only have we demonstrated a successful model for stealth CME
eruptions but we also argue that essentially all slow streamer blowout
CMEs are likely to be some variation of this scenario.

\citet{Lynch2010} presented a back of the envelope calculation of
the occurrence rate and total number of slow streamer blowout CMEs
generated by the coronal evolution in response to large-scale,
photospheric processes such as differential rotation.  We
obtained an estimate of approximately eight CMEs per month over the whole
streamer belt; this was approximately half the rate in CDAW LASCO
CME catalog for slow CMEs $>$30$^\circ$ in angular width during the
2008--2009 solar minimum. The starting point of this estimate was
generating the required magnetic fluxes from the initial PFSS helmet
streamer field to match the observed in situ flux content of the
6--7 June 2008 ICME magnetic cloud flux rope at STEREO-B.  Here our
MHD simulation results have confirmed the basic assumption underlying
the \citet{Lynch2010} discussion: a large-scale shear (equivalent
to $\sim$2~weeks of differential rotation over 40$^\circ$ latitude)
applied to the bipolar helmet streamer could produce a slow, flux
rope-like streamer blowout eruption, such as observed by STEREO-A
on 1--2 June 2008.  Our simulation results are
consistent with the \citet{HudsonLi2010} conclusion that ``the
observation that CMEs continue into solar minimum whereas flares
do not, or else become less important, suggests the existence of a
CME mechanism basically independent of the activity level.''

The ICME flux rope orientation over the solar cycle
\citep[e.g.,][]{Zhao1996,Bothmer1997,Bothmer1998,Mulligan1998,Li2011,Li2014}
provides additional evidence that global-scale photospheric
evolutionary processes, modulated by the magnetic field structure
and evolution of the corona, are the underlying cause for a significant
fraction of CME activity.  \citet{Lynch2005} examined the flux and
helicity content of magnetic cloud flux rope ICMEs from 1995 to 2003.
We showed the net cumulative helicity of 108 ``slow events'' ($V_{\rm
ICME} < 500$~km~s$^{-1}$) essentially averaged to zero over the
8 year period. A net cumulative helicity of zero is exactly as
expected if the ICME helicity originates from a large-scale velocity
pattern that is symmetric with respect to the equator acting on a
large-scale, antisymmetric magnetic field, i.e., differential
rotation acting on the global dipole magnetic field \citep{DeVore2000}.

We conclude from this discussion and from the results above that,
in many ways, stealth CMEs bridge the gap between the standard
events driven by filament channel eruption and the small-scale
plasmoids that are observed to be continuously emitted from streamer
tops \citep{Viall2015}. According to the S-Web model, these plasmoids
are due to the reconnection dynamics of coronal hole boundaries
driven by the photospherc convective motions \citep{Antiochos2011}.
If so, then fast CMEs, stealth CMEs and streamer plasmoids may well
constitute a continuum of manifestations of a single process: the
ejection of magnetic stress into the heliosphere via reconnection.
For fast CMEs the magnetic stress builds up on the lowest-lying
flux near polarity inversion lines. For streamer top plasmoids the
stress is injected on the highest-lying flux near the boundary of
the closed field system. In stealth CMEs the stress is injected at
intermediate heights by the large-scale differential rotation. In
all cases the stress appears to be released by the formation and
ejection of a reconnection plasmoid similar to the breakout CME
simulations of \citet{Karpen2012}.  If this ``unified'' scenario
is correct, it could greatly simplify our understanding and modeling
of coronal evolution. Detailed studies of stealth CMEs, which are
relatively easy to observe accurately due to their large spatial
and temporal scales, may well provide key insights into phenomena
at smaller and/or much faster scales.

\subsection{Implications for Space Weather Forecasting}
\label{ssxn.disc.spw}

The STEREO mission space weather capabilities and NOAA Space Weather
Prediction Center forecasting requirements are discussed by
\citet{Biesecker2008SSR}.  Typical on-disk signatures such as soft
X-ray sigmoid/arcade emission, filament eruptions in H$\alpha$ or
EUV, UV flare ribbon emission, and/or large-scale coronal dimmings,
followed by halo or partial-halo CMEs, act as a 2--5 day precursor
warning for potential geomagnetic storms from Earth-impacting events.
Stealth CMEs are a problematic type of event to effectively forecast
due to the lack of on-disk signatures and their halo signatures
being difficult to observe in coronagraph data.  For example,
\citet{Webb2000JGR} analyzed a faint Earth-directed halo CME that
caused significant geomagnetic activity.

The 1--2 June 2008 stealth CME produced an extremely faint halo CME
signature in the STB COR2 running difference images only visible
for a few frames \citep{Robbrecht2009}. However, as we have shown
here and in \citet{Lynch2010}, STA was able to image a well-resolved,
edge-on flux rope CME in COR2 and track its propagation through HI1
and HI2 all the way to its impact with STB at 1 AU.  This fortuitous
observing geometry was due to the $\approx$55$^\circ$ separation
between the STEREO spacecraft.  Our simulation's synthetic
running difference images from the STB viewing perspective do contain
halo CME-like signatures but only at the level of the running difference
fluctuations caused by the ambient solar wind outflow.  In other
words, it would be almost impossible to identify the simulated CME
from this viewpoint alone.

The L5 Lagrange point at 1 AU, located 60$^\circ$ from the Sun-Earth
line, is ideally suited for identifying and observing such
Earth-directed stealth CME events. Indeed, the prospect of an L5
space weather mission for operational forecasting has generated
considerable community and agency support
\citep[e.g.,][]{Gopalswamy2011JASTP, Howard2012AGUFM, Lavraud2014cosp,
Trichas2015Hipp}.  Our simulation results, the eruption of a
large-scale coherent flux rope CME towards STB with almost no
discernible halo or partial-halo CME signatures, highlights the
importance of supplementing the traditional L1 viewpoint for space
weather forecasting.

\section{Summary and Conclusions}
\label{sxn.concl}

We have presented a numerical simulation of the 1--2 June 2008 stealth
CME and shown that all of the qualitative large-scale properties
of the slow streamer blowout CME are reproduced by our calculation.

1. The 3-D preeruption global coronal structure in multiviewpoint
synthetic white-light coronagraph images captures the overall shape
and location of the helmet streamer and pseudostreamer structures
in CR2070.

2. We energized the bipolar helmet streamer over the source region
estimate by \citet{Robbrecht2009} via large-scale shearing flows
with respect to the streamer belt polarity inversion line, obtaining
a total foot point displacement equivalent to $\sim$2~weeks of
differential rotation.

3. The sheared helmet streamer arcade expands slowly over the
next several days, gradually opening more restraining overlying
flux into the solar wind. The gradual opening of the helmet streamer
flux facilitates continued arcade expansion, which opens more closed
flux in a positive feedback loop similar to the runaway expansion
in the \citet{Antiochos1999} breakout model for CME initiation.
However, unlike the breakout model, in which the restraining flux
is rapidly transferred through magnetic reconnection, the gradual
expansion-driven opening of helmet streamer flux into the solar
wind is a relatively slow disruption of pressure balance.

4. Given the low magnetic field strengths associated with the
quiet-Sun flux distribution and the spatial scale of the helmet
streamer system, the stored magnetic energy released during our
stealth CME eruptive flare reconnection is $\sim$10$^{30}$~erg over
a time of $\gtrsim$20 h.  Our estimate of the available magnetic
energy flux for observable flare heating and emission during the
eruption is 2 orders of magnitude below the energy flux required
to heat the ambient background corona.

5. Our simulation produces the observed white-light signatures
of the X point flare current sheet in the STA COR1 field of view,
the three-part structure of flux rope CMEs in synthetic running difference
images, and the height-time and velocity profiles of the CME
propagation through $\sim$15$R_\odot$ in excellent agreement with
the STA COR2 observations.  Our velocity profile implies that the
stealth CME structure is passively advected with the background
solar wind.

6. Our results support the \citet{Howard2013} argument that stealth
CMEs are not fundamentally different from most slow streamer blowout
CMEs; they simply represent the lowest-energy range of the (slow)
CME distribution.

We have discussed the implications of our stealth CME simulation
results for sheared arcade models of CME initiation, their relationship
to the global evolution of the large-scale corona, and for future
research on the space weather consequences of stealth CMEs.


%
%
%
%
%
%
%

\begin{acknowledgments}

B.J.L., Y.L., J.G.L., and G.H.F. acknowledge support from AFOSR YIP
FA9550-11-1-0048, NSF AGS-1249150, NASA HTP NNX11AJ65G, the Coronal
Global Evolutionary Model (CGEM) project NSF AGS-1321474, and NASA
STEREO Data Analysis funds.  S.M., C.R.D., and S.K.A. acknowledge
support from NASA HSR and LWS TRT programs.  The computational
resources supporting this work were provided by the NASA High-End
Computing Program through the NASA Center for Climate Simulation
at Goddard Space Flight Center.
SECCHI data used here were produced by an international consortium
of the Naval Research Laboratory (USA), Lockheed Martin Solar and
Astrophysics Lab (USA), NASA Goddard Space Flight Center (USA),
Rutherford Appleton Laboratory (UK), University of Birmingham (UK),
Max-Planck-Institut for Solar System Research (Germany), Centre
Spatiale de Li\'{e}ge (Belgium), Institut d'Optique Th\'{e}orique
et Applique\'{e} (France), and Institut d'Astrophysique Spatiale
(France). \url{https://stereo-ssc.nascom.nasa.gov/}.
The SOHO/LASCO data used here are produced by a consortium of the
Naval Research Laboratory (USA), Max-Planck-Institut for
Sonnensystemforschung (Germany), Laboratorie d'Astrophysique Marseille
(France), and the University of Birmingham (UK). SOHO is a project
of international cooperation between ESA and NASA (\url{http://lasco-www.nrl.navy.mil/}).
The MK4 coronagraph data are provided courtesy of the Mauna Loa
Solar Observatory, operated by the High Altitude Observatory, as
part of the National Center for Atmospheric Research (NCAR). NCAR
is supported by the National Science Foundation (\url{https://www2.hao.ucar.edu/mlso/mlso-home-page}).

\end{acknowledgments}

%
%


\end{document}